\documentclass[twocolumn,tighten,numberedappendix]{aastex62}
\usepackage{lineno}
\usepackage{bm}
\usepackage{xspace}
\usepackage{enumitem}
\usepackage{graphicx}
\usepackage{color}
\usepackage{comment}
\usepackage{amsmath}
\usepackage[T1]{fontenc}

\providecommand{\dodoi}[1]{doi:~\href{http://doi.org/#1}{\nolinkurl{#1}}}
\providecommand{\doarXiv}[1]{arXiv:~\href{https://arxiv.org/abs/#1}{\nolinkurl{#1}}}

\def\src{3XMM~J215022.4$-$055108}
\def\xmm{XMM-{\it Newton}}
\def\chan{{\it Chandra}}
\graphicspath{{./}{figures/}}

\shorttitle{The Black Hole Mass and Spin of J215022.4-055108}
\shortauthors{Wen et al.}
\begin{document}

\title{Mass, Spin, and Ultralight Boson Constraints from the Intermediate Mass Black Hole in the Tidal Disruption Event 3XMM~J215022.4$-$055108}
\correspondingauthor{Sixiang Wen}
\email{wensx@email.arizona.edu}
\author[0000-0002-0934-2686]{Sixiang Wen}
\affiliation{University of Arizona, 933 N. Cherry Ave., Tucson, AZ  85721}

\author[0000-0001-5679-0695]{Peter G.~Jonker}
\affiliation{Department of Astrophysics/IMAPP, Radboud University, P.O.~Box 9010, 6500 GL, Nijmegen, The Netherlands}
\affiliation{SRON, Netherlands Institute for Space Research, Sorbonnelaan 2, 3584~CA, Utrecht, The Netherlands}
\author{Nicholas C. Stone}
\affiliation{Racah Institute of Physics, The Hebrew University, Jerusalem, 91904, Israel}

\author[0000-0001-6047-8469]{Ann I. Zabludoff}
\affiliation{University of Arizona, 933 N. Cherry Ave., Tucson, AZ  85721}

\begin{abstract}
We simultaneously and successfully fit the multi-epoch X-ray spectra of the tidal disruption event (TDE) 3XMM~J215022.4$-$055108 using a modified version of our relativistic slim disk model that now accounts for angular momentum losses from radiation. We explore the effects of different disk properties and of uncertainties in the spectral hardening factor $f_c$ and redshift $z$ on the estimation of the 
black hole mass $M_\bullet$ and spin $a_\bullet$.
Across all choices of theoretical priors, we constrain $M_\bullet$ to less than
2.2$\times 10^4$ $M_\odot$
at $1\sigma$ confidence. Assuming that the TDE host is a star cluster associated with the adjacent, brighter, barred lenticular galaxy at $z=0.055$, we constrain $M_\bullet$ and $a_\bullet$ to be $1.75^{+0.45}_{-0.05}\times 10^4$ $M_\odot$ and $0.8^{+0.12}_{-0.02}$, respectively, at $1\sigma$ confidence. 
The high, but sub-extremal, spin suggests that, if this intermediate mass black hole (IMBH) has grown significantly since formation, it has acquired its last e-fold in mass in a way incompatible with both the “standard” and “chaotic” limits of gas accretion.  
Ours is the first clear IMBH with a spin measurement.  
As such, this object represents a novel laboratory for astroparticle physics; its $M_\bullet$ and $a_\bullet$ place tight limits on the existence of ultralight bosons, ruling out those with masses $\sim$10$^{-15}$ to $10^{-16}$ eV. 
\end{abstract}


\section{Introduction}
The electromagnetic flares associated with tidal disruption events (TDEs), in which a star is broken apart by the differential gravity of a supermassive black hole \citep[SMBH;][]{Hills75, Lidskii79, Rees1988, Evans1989}, offer a direct and promising route to constrain the mass and spin of otherwise dormant SMBHs.  TDE candidates were first found in soft X-ray wavelengths \citep{Bade+96, Komossa+04}, where emission is often but not always quasi-thermal \citep{Saxton+17}.  TDEs have also been discovered in thermal optical/UV radiation \citep{Gezari+06, van+11}; optical surveys dominate the current rate of TDE detection, finding $\mathcal{O}(10)$ events per year \citep{vanVelzen+20}.  In principle, emission at all of these wavelengths can be used to constrain the handful of underlying event parameters, such as SMBH mass $M_\bullet$ and spin $a_\bullet$.

TDEs have two advantages over traditional techniques for measuring $M_\bullet$ and $a_\bullet$. First, in contrast to other methods, TDEs probe lower mass SMBHs, and possibly even intermediate mass black holes (IMBHs). Loss cone modeling predicts that the volumetric rate of TDEs is dominated by the lowest mass dwarf galaxies with a high black hole (BH) occupation fraction \citep{WangMerritt04, StoneMetzger16}.  A volume-complete TDE sample would therefore measure the low end of the SMBH mass function.  The gravitational influence radius of IMBHs is usually too compact for stellar dynamical mass measurements\footnote{IMBH candidates can be found dynamically \citep[e.g., ][]{Nguyen+19}, but only in the very nearest galaxies.}, and the faintness and short light-crossing times of AGN in dwarf galaxies have limited reverberation mapping mass measurements to a handful of systems \citep[see, e.g., ][for a review]{Reines+13,Greene+20}.  

The second advantage TDEs bring to the black hole census is their unique ability to measure $a_\bullet$ in {\it quiescent} galactic nuclei.  Far from the horizon, $a_\bullet$ is a higher order correction to the gravitational potential, challenging to observe even for resolved S-star orbits in the Milky Way \citep{Merritt+10}, and impossible for unresolved stellar orbits around extragalactic SMBHs.  Iron K-$\alpha$ spectroscopy allows spin measurements in bright AGN \citep{Reynolds20}, which, however, may not be representative of the SMBH population \citep{BertiVolonteri08}.  

The masses of the black holes causing TDEs are typically inferred from galaxy scaling relations, such as the $M_\bullet-\sigma$ or $M_\bullet-M_{\rm bulge}$ correlations \citep{Wevers+17, Wevers+19}, an indirect approach that is not well-calibrated for IMBHs.  It is therefore desirable to measure $M_\bullet$ directly, from the light curves or spectra of the flares themselves.  Multiple models exist for constraining $M_\bullet$ from a TDE optical/UV light curve \citep{Guillochon+14, Mockler+19, Ryu+20}, but challenges for this approach include the unknown power source \citep{Loeb+1997, Piran+15, MetzgerStone16} and three-dimensional geometry \citep{Guillochon+14, Shiokawa+15, Dai+18} of the optical/UV photosphere.  Likewise, it is not clear how (or even whether) $a_\bullet$ will affect the optical/UV emission\footnote{One possibility, not yet quantified in light curve modeling, is the sub-leading role spin plays in setting the self-intersection radii of eccentric debris streams \citep{Wevers+17}.}. Neither scaling relations nor optical/UV light curve fitting have produced constraints on $a_\bullet$.

These limitations motivated us to use X-ray continuum fitting to determine $M_\bullet$ and $a_\bullet$ in TDEs.  To do so, we extended stationary general relativistic “slim disk” accretion models from stellar-mass black holes to SMBHs for the first time \citep[][hereafter W20]{Wen+20}. These slim disk models extend standard thin disk accretion theory \citep{ShakuraSunyaev1973,NovikovThorne1973} to accretion rates 
comparable to or larger than approximately ten per cent of the Eddington limit, where sub-Keplerian gas motion and advective heat losses can no longer be neglected \citep{Abramowicz+1988}.  We ray-traced the trajectories of photons from the image plane to the disk surface, including gravitational redshift, Doppler, and lensing effects self-consistently. 

In W20, we applied these general relativistic slim disk models to two well-studied SMBH TDEs: ASASSN-14li \citep{Holoien+16a} and ASASSN-15oi \citep{Holoien+16b}, placing strong constraints on $M_\bullet$ for both flares and on $a_\bullet$ for ASASSN-14li.  In this paper, we apply our models to X-ray observations of \src, hereafter ``J2150.'' The J2150 flare is a luminous X-ray outburst in a small optical source 
adjacent to the large, barred lenticular galaxy 6dFGS~gJ215022.2$-$055059 at $z=0.055$ \citep[][hereafter L18]{Lin+18}. J2150, one of the most compelling IMBH TDE candidates to date, was first detected by L18 in the \xmm\, X-ray source catalog. If the position is not a chance association, the host is a star cluster 
of mass $\sim 10^7 M_\odot$ and a half-light radius of about 27 pc, offset by $\approx 12$ kpc from the lenticular's center  \citep[][hereafter L20]{lin+20}. L18 fit standard thin disk accretion models to this flare and estimated an IMBH mass of $5\times 10^4 M_\odot \lesssim M_\bullet \lesssim 1\times 10^5 M_\odot$.  Their work hints that $a_\bullet$ is large, but they only consider two possible values for it in their fits. In this paper, we re-reduce and reanalyze existing multi-epoch X-ray observations and simultaneously fit the continuum with our relativistic slim disk model to constrain $M_\bullet$ and $a_\bullet$.


In \S \ref{sec:method}, we summarize our theoretical model for quasi-thermal X-ray emission from TDE accretion disks.  In \S \ref{sec:data}, we review the X-ray observations of J2150 and our data reduction methods.  In \S \ref{sec:results}, we present our X-ray spectral fits, the resulting constraints on the ($M_\bullet, a_\bullet$) plane, sources of uncertainty, and implications for particle physics.  We summarize in \S \ref{sec:conclusions}.  Throughout, we assume a flat cosmology with $H_0 = 69.6~\rm{km}~\rm{s}^{-1}\rm{Mpc}^{-1}$, $\Omega_M=0.29$ and $\Omega_\Lambda=0.71$.

\section{METHODOLOGY}
\label{sec:method}

In this work, we follow the general procedures of W20. We use the general relativistic stationary slim disk to model the dynamic TDE accretion disk, a color-modified, multi-color black body model \citep{DE18} to calculate the local X-ray emission, and a geodesic ray-tracing code \citep{JP11}, which includes gravitational redshift, Doppler, and lensing effects self-consistently, to calculate the synthetic X-ray spectrum.  

For high spin BHs, the radiative efficiency $\eta$ can be up to 0.42, which would allow the radiation to take away a significant part of the angular momentum \citep{Abramowicz+1996}. Here, we update our code to account for the angular momentum loss by radiation in the disk equations. For the reader’s convenience, we write the underlying slim disk equations in the Appendix \ref{app:slimdisk} and explore the importance of angular momentum loss by radiation.  Neglecting this angular momentum loss produces changes in the local effective temperature that are always $<10\%$, and usually $<5\%$ (see Fig. 8 in Appendix \ref{app:slimdisk}).  The effect is maximized for black holes with spins near the Thorne limit ($a_\bullet = 0.998$; \citealt{Thorne74}) and sub-Eddington accretion rates.

A fully circularized TDE debris stream would form an accretion disk with an initial radius $R_{\rm c}=\frac{2R_{\rm t}}{\beta}$, where the tidal disruption radius is
\begin{align}
    R_{\rm t} =& R_\star \left( \frac{M_\bullet}{M_\star} \right)^{1/3}\\ \notag \approx& 1.0 \times 10^3 R_{\rm g} \left(\frac{M_\bullet}{10^4 M_\odot} \right)^{-2/3} \left(\frac{M_\star}{M_\odot} \right)^{-1/3} \left(\frac{R_\star}{R_\odot} \right),
\end{align}
to within factors of order two \citep{Guillochon13}.  Here $\beta = \frac{R_{\rm t}}{R_{\rm p}}$ with $R_{\rm p}$ the periastron radius of the star's orbit, $R_{\rm g} = GM_\bullet / c^2$ is the gravitational radius, and $M_\star$ and $R_\star$ are the mass and radius, respectively, of the disrupted star.

As twice the tidal radius, in an IMBH system, is about one thousand gravitational radii away from the IMBH (and thus too cold to produce significant X-ray flux), we set the outer edge of the disk to $\le$ 600 gravitational radii. The error on the flux caused by the choice of outer edge is $< 1\%$ (see appendix \ref{rout_rin}). However, the choice speeds up our calculations by at least 4 times. When we do the ray tracing, we cut off the disk at the innermost stable circular orbital (ISCO), due to a singularity inside the ISCO when calculating the spectral hardening factor. The error on the flux caused by the different choices of inner edge is less than $2\%$ (see appendix \ref{rout_rin}). For an accretion disk around a high-spin BH, the error is less than $0.5\%$ (see Appendix \ref{rout_rin}).    

We do not assume any prior on the disk accretion rate from TDE gas fallback hydrodynamic simulations; instead, we treat the disk accretion rate as a free parameter for the fit in each epoch of data. The free parameters for the slim disk model are $M_\bullet$, $a_\bullet$, $\dot m_i$, and $\theta$. Here, the subscript index $i$ denotes the $i$-th observational epoch and $\theta$ is the inclination angle of the accretion disk with respect to our line of sight. The X-ray spectrum is subject to circumnuclear and interstellar extinction, so we introduce the extinction parameter $N_{\rm H,i}$. 


The three largest assumptions in our model, as applied to TDEs, are that (1) the dynamic inner disk can be approximated by a time series of stationary accretion disks, (2) the inner accretion disk is axisymmetric, and (3) the inner disk is always aligned with the BH equatorial plane. The first assumption is true at all times, because, at each timestep, the accretion rate across the disk is nearly the same.  At early times, this is because the viscous timescale is much shorter than the mass fallback timescale. 
At late times, this is because the decreased importance of mass fallback has caused the disk to settle into a self-similar spreading state, where mass accretion rates are almost constant in the inner disk, the source of the X-ray emission fit with our model.

The latter two assumptions are questionable in the earliest phases of TDEs, but are likely better for later observational epochs.  Initially, the gas from the disrupted star returns to the BH on highly eccentric trajectories, which dissipate excess kinetic energy in shocks, thereby circularizing into an accretion disk.  The efficiency and progression of this circularization process is currently unknown for realistic TDE parameters, and if the inner disk remains significantly non-axisymmetric, it will bias the results of our continuum fitting\footnote{Note that the slim disk models from W20 account self-consistently for sub-Keplerian fluid motion, one aspect of incompletely circularized accretion flows, but not for the large-scale apsidal misalignment characteristic of a globally eccentric flow.}.  However, we note that for IMBHs, the characteristic circularization radius $R_{\rm c} \sim 1000 R_{\rm g}$ is about two orders of magnitude larger than the radii that emit most of the observed X-ray radiation, and matter that has reached scales of $\sim R_{\rm g}$ has thus dissipated about two orders of magnitude in orbital energy through shocks or magnetized turbulence.  Even if early-time epochs in TDE disks feature globally eccentric structures, it seems reasonable to assume -- especially in the case of IMBHs -- that the X-ray emitting inner annuli will have mostly circularized.

Likewise, orbital dynamics of the loss cone suggest that disrupted stars approach BHs from a quasi-isotropic distribution of directions, implying that TDE disks should be born with a substantial tilt \citep{StoneLoeb12} that is not accounted for in our models.  The most immediate observational consequence of disk tilt will be a softening of the X-ray spectrum (as the tilted analogue of the ISCO sits further out than the ISCO itself); there may also be more complex lensing effects related to the geometry of null geodesics in the axisymmetric Kerr spacetime.  This tilt will decay over time due to the onset of the Bardeen-Petterson effect \citep{StoneLoeb12}, inter-annulus torques in a globally precessing thick disk \citep{Franchini+16}, and torques from returning debris streams impacting a misaligned, precessing accretion flow \citep{XiangGruess+16, ZanazziLai19}.  Because this effect may potentially bias the results of early-time observations, we will perform two different multi-epoch fits: one that uses all observational epochs, and one which only employs the late-time epochs, when we can be more confident that the inner disk has aligned itself (this alternate fit focused on late epochs may also be more trustworthy with regards to the assumption of axisymmetry).  

\section{Data Reduction}
\label{sec:data}
We will include in our analysis pointed observations obtained with both the \xmm\, \citep{Jansen+01} and the \chan\, \citep{weisskopf+02} satellites. Table~\ref{tab:xmmchan} lists some properties of these observations. Note that all these observations were also used by L20 but our analysis differs from theirs in important points. First, we employ Poisson statistics in all of our spectral fits \citep{Cash1979}. As shown in the work of \citet{Kaastra17}, the use of $\chi^2$ statistics in spectral fits even for 20-30 counts per spectral bin will bias the results of the fit. Second, we correct for the influence of an interloping nearby source on the \xmm\, spectral fits (see \ref{sec:chan} below).

\subsection{\chan\,data}
\label{sec:chan}
Two \chan\, observations of J2150 exist (see Table~\ref{tab:xmmchan} for more information). We used {\sc ciao} version 4.12 for our \chan\, data analysis (\citealt{Fruscione+06}). During the first observation obtained in 2006, the source was observed serendipitously on the ACIS-I CCD array and at a large off-axis angle where the \chan\, point spread function is degraded with respect to that on-axis.  We extracted the source and background spectrum using circular regions with a radius of 7.7 and 60\arcsec, respectively.

\begin{figure}
\plotone{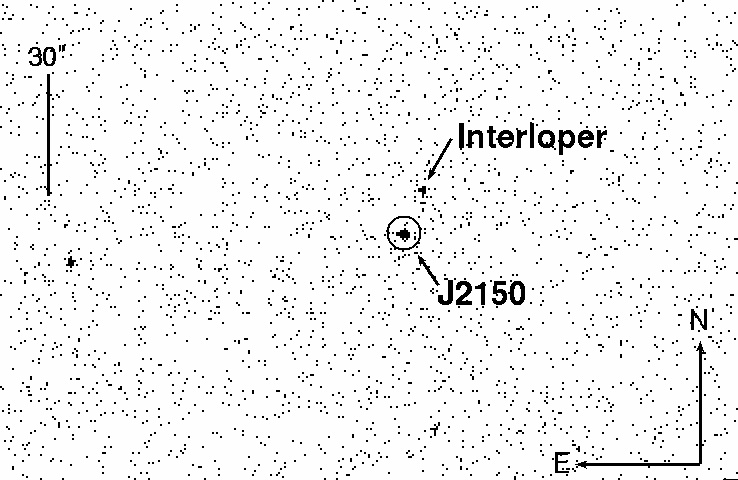}
\caption{\chan\, 0.1--7 keV image from observation ID 17862 of the field around J2150 showing the nearby interloper that contaminates the \xmm\, spectra of J2150. Note that the circle centered on the J2150 position has a radius of 1 pn pixel (i.e., 4.1\arcsec). In our spectral fits of the \xmm\, data we treated this interloper as an additional contribution to the background, assuming its flux and spectral shape did not vary over time. 
}
\label{J2150chan}
\end{figure}

During the on-axis \chan\, observation of 2016 with the ACIS-S3 CCD, in addition to J2150, another source was clearly detected at $\alpha$: 21:50:22.2, $\delta$: 05:50:58.7 (J2000), a location consistent with the nucleus of the nearby, barred lenticular galaxy (6dFGS~gJ215022.2$-$055059; see L18 and L20; see our Fig.~\ref{J2150chan}). This location is 11.2\arcsec\, away from the position of J2150. As this source position is close enough, given the size of the \xmm\, point spread function, to influence the \xmm\, TDE spectra, we assessed its flux and spectral shape to treat it as background in the spectral fitting of the \xmm\, data (assuming its flux and spectrum are constant). 
We extracted the spectrum of both this interloper and J2150 using a circular region with a radius of 1.5\arcsec. We used a source-free, nearby circular region with a 1\arcmin\, radius on the same detector to estimate the background spectrum. For all our fits, we first fit the background spectrum separately with two power law models. Next, to correct the TDE spectrum for the background contribution, the best model-fit values for the background model are kept fixed during the fit of the TDE spectrum.

\subsection{\xmm\,data}
\label{sec:xmm}
We run the {\sc SAS} v18 (20190531) tools under the HEASOFT {\sl ftools} software version 6.26.1 to extract the spectra of J2150 and filter both the EPIC pn as well as the MOS detector data. All the observations are done with the pn and the MOS detectors in Prime Full Window mode, providing a time resolution of 73.4~ms and 2.6~s for the pn and MOS detectors, respectively. We filtered the pn and MOS data for periods of enhanced background radiation, where we require that the 10--12 keV detection rate of pattern 0 events is $<$~0.4 counts s$^{-1}$ for the pn and that the $>$~10 keV detection rate of pattern 0 events is $<$~0.35 counts s$^{-1}$, for both MOS detectors. The effective exposure time for each observation after filtering is given in Table~\ref{tab:xmmchan}. To investigate if pile-up is important, we use the {\sc SAS} command {\sc epatplot} to compare the observed and expected number of single and double event pattern as a function of photon energy. We conclude that pile-up is not important.

We extracted the spectrum of J2150 using a circular aperture of 30\arcsec\, radius centered on the optical position of J2150 for the pn and the both MOS detectors, except for the pn data of observation ID 0603590101, where we used a radius of 20\arcsec\, to avoid bad pixels falling into the source extraction region. To extract the background spectrum, we used a circular region on the same CCD as close to the position of J2150 as possible with a radius of 75\arcsec, except for the pn observations of observation ID 0603590101 and 0823360101, where we had to use a circle with radius of 45\arcsec\, to avoid bad pixels falling into the extraction region.


\begin{deluxetable*}{lrcc}
\tablecaption{XMM--{$\it Newton$} (top part of the Table) and \chan\, (bottom) observations of \src\, used in this paper.}
\tablewidth{0pt}
\tablehead{
\colhead{Observing } & \colhead{Start date \& time} & \colhead{Exp time } & \colhead{Counts$^{\dagger}$} \\
\colhead{ID } & \colhead{[UTC]} & \colhead{pn/MOS1/MOS2 [ks]} & \colhead{pn/MOS1/MOS2} 
}
\startdata
0404190101 & 2006-05-05 12:24:35 & 22.2/49.6/49.6 & 2605/3559/3000   \\
0603590101 & 2009-06-07 07:53:31 & 40.8/68.3/67.9 & 3724/1687/1629    \\
0823360101 & 2018-05-24 08:28:11 & 40.3/57.8/57.8 & 1179/327/362   \\
\hline
6791 & 2006-09-28 20:49:55 & 100.6 & 4801   \\
17862 & 2016-09-14 07:31:09 & 77.1 &  152
\label{tab:xmmchan}
\enddata
\tablecomments{$^{\dagger}$ Counts detected after filtering on photon energy between 0.3--10 keV for the pn, MOS1, MOS2 detectors and 0.3--7 keV for the \chan\, ACIS-I and -S detectors for observation IDs 6791 and 17862, respectively. }
\end{deluxetable*}

\section{Results and Discussion}
\label{sec:results}

Throughout this paper, we fit the spectra by using {\sc XSPEC} version 12.11.1 \citep{Arnaud1996} applying Poisson statistics (\citealt{Cash1979}; {\sc C-stat} in {\sc XSPEC}). We quote all the parameter errors at a $1\sigma$ $(68.3\%)$ confidence level (CL), using the method $\rm{Statistic} = \rm{Statistic_{best-fit}} + \Delta\,C$ \citep{Arnaud1996} and assuming 
$\Delta Cstat = 1.0$ and $\Delta Cstat = 2.3$ for single and two parameter models, respectively \citep{Mao+18}. We explore the statistical properties of the fitting results further in Appendix~\ref{cstatisitc}.

\subsection{Model Fitting}
We fitted the \chan\, observation ID 17862 spectrum of the interloper 
and found it to be well-fit ($Cstat/\nu=42.22/38$) with a power law with index 1.7. 
We add this power law as an additional background component to our \xmm\, spectral fits of J2150. 

The quasi-thermal slim disk model of the previous section forms the basis for our X-ray spectral fitting and parameter estimation for J2150. We fit the multi-epoch spectra of J2150 by combining our slim disk model with two absorption parameters, $N_{\rm H}$, one fixed at $2.6\times 10^{20}\,\rm{cm}^{-2}$ at redshift $z=0$ (L18) to account for Galactic absorption, the other allowed to float using a redshift fixed at $z=0.055$ to describe the effect of any extinction in the host, TDE, and perhaps the nearby, barred lenticular (although TDE may be in front of it). The general absorption models  {\sc phabs} and {\sc zphabs} in XSPEC \citep{Arnaud1996} are added as multiplication models to our slim disk model to account for these components. 

Using Cash statistics \citep{Cash1979}, we need to fit the source and background together. As explained above, the background spectrum in the \chan\, observations is well-fit using two power law models. The \xmm\, background can be well fitted with two power laws plus two Gaussian emission lines (arising from the satellite, and there is only one Gaussian emission line for the pn spectra). If we would ignore the two emission lines at about 1.5 and 1.8 keV, the results would be biased favoring high spin values. During the fit of J2150, we fix the respective background models at their best-fit values from the background-only fit to accelerate the calculation. The final fit-function is a combination of the background fit-function plus our absorbed slim disk model. The complete fit-function in XSPEC 
is {\sc po+po+po+agauss+agauss+phabs( zphabs(slimdisk))}, with the first {\sc po} accounting for the interloper contamination. However, for the \chan ~spectra, which are neither affected by the interloper nor by the background Gaussian emission lines, we fix the normalization of the interloper power law and the two Gaussian emission lines to 0.

Following the procedure of W20, we fit the five spectra simultaneously, by fixing ($M_\bullet$, $a_\bullet$) at the grid value, while allowing all five accretion rates ($\dot m_i$), all five absorption parameters ($N_{\rm H,i}$), and the one inclination ($\theta$) to float. In order to evaluate the significance of each ($M_\bullet$, $a_\bullet$) pair, we minimize $Cstat$ for each ($M_\bullet, a_\bullet$) in our ($M_\bullet$, $a_\bullet$) grid. The parameter priors are listed in Table 2.
In the initial fit, we use a spectral hardening factor $f_{\rm c}$ as calculated by \citep{DE18} (see W20 for more detail; we call this the fiducial treatment of $f_{\rm c}$, with more details in Appendix \ref{f-fc}). In all the fitting, we allow for the possibility that the absorption local to the TDE changes with time by keeping the host$+$TDE absorption component $N_{\rm{H,i}}$ as a free parameter, while the absorption from Milky Way has been fixed at $2.6\times 10^{20} ~\rm{cm}^{-2}$ (L18). 

\begin{deluxetable}{ccccccc}
\tablecaption{Parameter Priors for Fitting X-Ray Spectra.}
\tablewidth{0pt}
\tablehead{
\colhead{$M_\bullet^a$} & \colhead{$a_\bullet^a$} & \colhead{$\dot m_i^b$} & \colhead{$\theta$} & \colhead{$N_{\rm H, i}$} &  \\
\colhead{$[10^4 M_\odot]$} & \colhead{} &  \colhead{$[\rm Edd]$} & \colhead{$[^{\circ}]$} & \colhead{$[10^{21}{\rm cm}^{-2}]$} & 
}
\startdata
 [0.15, 10] & [-0.9, 1.0)  & [0.05, 100] & [5, 90] & (0, 1] &  \\
\enddata
\label{tab:prior}
\tablecomments{$^a$For individual epoch fits, $M_\bullet$ and $a_\bullet$ are discretely sampled at each grid point across the given ranges.~$^b$We use linear interpolation to estimate spectra for accretion rate values $\dot{m_i}$ between discrete grid points.
The accretion rate 
is calculated by assuming $\eta=0.1$ and listed in dimensionless Eddington units. }
\end{deluxetable}

In preliminary initial fits where we did not require $M_\bullet$ and $a_\bullet$ to be the same across all five epochs, we found that the last three of the five epochs can be well fitted and constrain the $M_\bullet$ and $a_\bullet$ to values that are mutually consistent within errors. However, the first two epochs only find their best fit at a high accretion rate and an extremely high $a_\bullet$. The constraint on $a_\bullet$ derived using the first two epochs is $> 3\sigma$ away from the value derived using the last three epochs. This inconsistency stems from the fact that, except for extremely high values for $a_\bullet$, the disk is not bright enough to fit the first two epochs under the initial assumptions.

This result suggests that the source was accreting at a highly super-Eddington rate during the first two epochs, as is theoretically predicted for main sequence-IMBH disruptions \citep[e.g.,~][]{Rees1988, Chen+18}. For such a highly super-Eddington accretion rate, the X-ray luminosity of the slim disk is virtually insensitive to the actual super-Eddington accretion rate value, but it does depend on the choice of $f_{\rm c}$ (W20). We provide more information on the role of $f_{\rm c}$ in this regime in Appendix \ref{f-fc}, but note here that the fiducial prescription of \citet{DE18} is only tailored for sub-Eddington accretion disks.

There are two possible reasons for why the disk is not bright enough during the first two epochs: (1) the fiducial $f_{\rm c}$ value we assume is too low; (2) the redshift $z$ to J2150 is overestimated. With these options in mind, we refit the data with two different models: Model 1, where we adopt the same fiducial $f_{\rm c}$ prescription but allow it to float for highly super-Eddington (i.e., the first two) epochs; Model 2, where we keep the fiducial $f_{\rm c}$ prescription, but allow the redshift of the source $z$ to float.

\begin{figure*}
\plotone{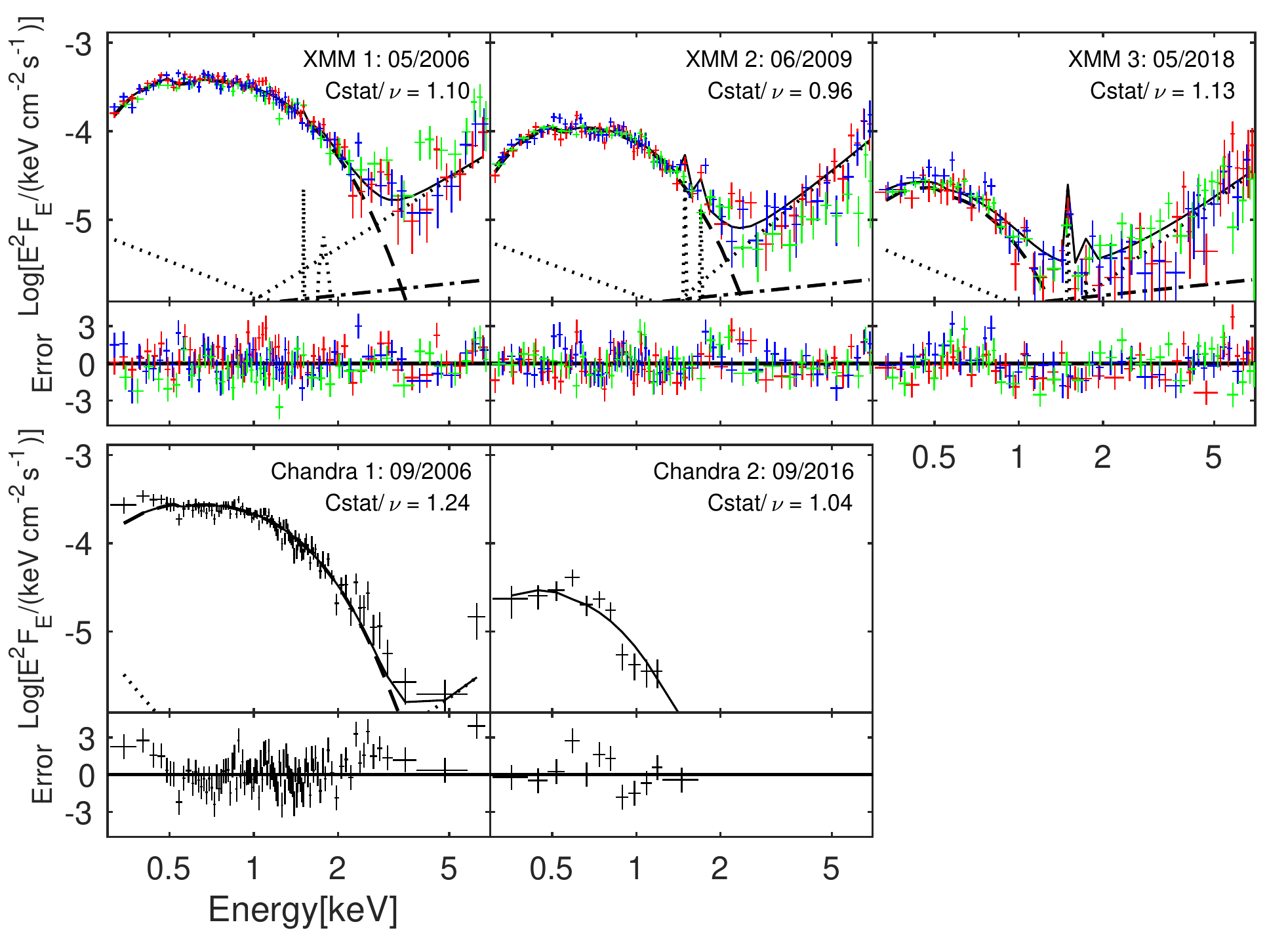}
\caption{Simultaneous slim disk fits to three XMM--{\it Newton} and two \chan\ spectra for J2150 from early (Epoch 1) to late (Epoch 5) times over a twelve-year period. The spectra are from the XMM--{\it Newton} PN (green line), MOS1 (red line) and MOS2 (blue line) detectors, which are most sensitive over the 0.3--10 keV range, and the \chan\, ACIS-I and -S detectors, which are most sensitive over the 0.3--7 keV range. All spectra include the background spectral contribution. The data are binned so that there is at least one X-ray photon per energy bin. These two features of the data allow us to employ Cash statistics \citep{Cash1979} in fitting the spectra. To avoid the plot becoming too crowded we show the Model~1 fit to only the MOS1 data for the top three panels, excluding the pn and MOS2 data. Each panel shows the best-fit slim disk (dashed line), the interloper power law (dot dashed line), the background (power law + power law + agauss + agauss) (dotted line), and the combined model (solid line). The interloper power law index and normalization are fixed to the same values for all the \xmm\, epochs derived from the fit to the interloper spectrum in the \chan\, data from observation ID 17862. The background spectral parameters are also fixed to the best-fit values derived from fits to the background spectra separately (see text). The accretion rate is allowed to float between the epochs. The sub-plot panels with Y-axis label ``Error" denote $\rm{(model - data)/ \sigma}$ for each spectral energy bin.
We provide the best-fit results in Table~\ref{J2150p}. }
\label{J2150sp}
\end{figure*}

\begin{deluxetable*}{cccccccccccccc}
\tablecaption{The best-fit parameters to J2150's five epochs of \xmm\, and \chan\, spectroscopic data derived using three different models.}
\tablewidth{0pt}
\tablehead{
\colhead{ }&& \colhead{XMM 1}  & & \colhead{Chandra 1} & & \colhead{XMM 2}  & & \colhead{Chandra 2} & & \colhead{XMM 3} \\ 
\colhead{Date } && \colhead{2006-05-05}  & & \colhead{2006-09-28} & & \colhead{2009-06-07}  & & \colhead{2016-09-14} & & \colhead{2018-05-24}  
}
\startdata
Model 1: flexible $f_{\rm c}$\\
$N_{\rm H}$ $[10^{20}{\rm cm}^{-2}]$ && $3.6\pm1.9$ && $0^{+0.7}$ && $3.8\pm0.5$ && $0^{+0.7}$ &&$1.3\pm0.9$\\
$f_{\rm c}$  && $2.4_{-0.06}^{+0}$ && $=f_{\rm c1}$ && - && - &&-\\
$\theta$ $ [{^\circ}]$ && $5.0_{-0}^{+10}$ && $=\theta_1$ && $=\theta_1$ &&$=\theta_1$ &&$=\theta_1$ &&\\
$\dot m^a$ ${\rm [Edd]}$ && $56\pm20$ && $5.1\pm1.0$ && $1.8\pm0.1$ && $0.35\pm0.02$ && $0.31\pm0.02$\\ 
$M_\bullet$ $[10^4{\rm M}_\odot]$ &&$1.75$&&$=M_{\bullet,1}$&&$=M_{\bullet,1}$&&$=M_{\bullet,1}$&&$=M_{\bullet,1}$\\
$a_\bullet$  &&$0.8$&&$=a_{\bullet, 1}$&&$=a_{\bullet, 1}$&&$=a_{\bullet, 1}$&&$=a_{\bullet, 1}$\\
$Cstat/\nu$ &&$298.94/270$ && $224.07/180$ && $271.17/283$ && $60.23/58$ &&$272.16/240$\\
\hline
Model 2: free $z$\\
$N_{\rm H}$ $[10^{20}{\rm cm}^{-2}]$ && $1.6\pm1.4$ && $0.3^{+1.0}_{-0.3}$ && $2.1\pm0.4$ && $0^{+0.7}$ &&$1.5\pm0.9$\\
$z$  && $0.017\pm0.004$ && $=z_1$ && $=z_1$ && $=z_1$ &&$=z_1$\\
$\theta$ $ [{^\circ}]$ && $61\pm26$ && $=\theta_1$ && $=\theta_1$ &&$=\theta_1$ &&$=\theta_1$ \\
$\dot m^a$ ${\rm [Edd]}$ && $37\pm28$ && $10.0\pm1.9$ && $2.9\pm0.2$ && $0.72\pm0.05$ && $0.64\pm0.04$\\ 
$M_\bullet$ $[10^4{\rm M}_\odot]$ &&$0.3$&&$=M_{\bullet,1}$&&$=M_{\bullet,1}$&&$=M_{\bullet,1}$&&$=M_{\bullet,1}$\\
$a_\bullet$  &&$-0.7$&&$=a_{\bullet, 1}$&&$=a_{\bullet, 1}$&&$=a_{\bullet, 1}$&&$=a_{\bullet, 1}$\\
$Cstat/\nu$ &&$301.20/270$ && $227.31/180$ && $270.43/283$ && $59.55/58$ &&$269.68/240$\\
\hline
Model 3: diskbb\\
$N_{\rm H}$ $[10^{20}{\rm cm}^{-2}]$ && $0.0^{+0.2}$ &&   $0^{+0.001}$        &&$2.1\pm 0.9$ && $0.0^{+7.2}$ &&$17.2\pm 5.2$  \\
$T_{\rm disk}[{\rm keV}]$ &&$0.261\pm0.003$ &&  $0.262\pm0.003$                       &&$0.215\pm0.005$ && $0.13\pm0.01$  &&$0.103\pm0.008$\\
$N_{\rm disk}^b$ && $14.4\pm 0.8$ &&    $9.6\pm 0.7$                                   &&$10.3\pm 1.5$ &&  $23\pm 9$ &&$271\pm 170$ \\  
$Cstat/\nu$ && $302.77/273$ && $236.71/179$ &&$270.4/282$ &&  $58.11/57$ && $254.87/239$ \\ 
\enddata
\tablecomments{For the slim disk model, the errors on the parameters are calculated keeping $M_\bullet$ and $a_\bullet$ fixed at their best-fit values. The total $Cstat/\nu$ for the combined five-epoch fit is $1126.58/1031=1.093$ and $1128.09/1031=1.094$, for Model 1 and Model 2, respectively. For the {\sc diskbb} model, the total $Cstat/\nu$ is $1122.86/1030=1.088$. The total $Cstat$ of the slim disk fit is close to the expected $Cstat=1124.1\pm1.4$ (\citealt{Kaastra17}), indicating a good fit to the data.  $^{\rm a}$ Accretion rate (in dimensionless Eddington units) is calculated by assuming a radiative efficiency of $\eta=0.1$. $^{\rm b}N_{\rm disk}$ is defined as $(R_{\rm in}/D_{10})^2\cos\theta$, where $R_{\rm in}$ is the inner disk radius in km, $D_{10}$ is the distance to the source in units of 10 kpc, and $\theta$ is the inclination angle of the disk.
In the fit to the \xmm\, spectra, we add a power-law component, $\Gamma=1.7$ and $A_{\rm pl}=1.1\times10^{-6}$ $\rm {photons ~s^{-1}~cm^{-2}~keV^{-1}}$, to account for the  contamination from the interloper which is the nuclear source of the nearby brighter barred lenticular galaxy. The Milky Way $N_{\rm H}$ absorption is fixed at $2.6\times10^{20}$ ${\rm cm}^{-2}$ (L18). The $N_{\rm H}$ shown in the Table is for the TDE, its host system, and possibly the nearby, interloping lenticular galaxy at $z=0.055$.}
\label{J2150p}
\end{deluxetable*}

\begin{figure*}[ht!]
\gridline{ \fig{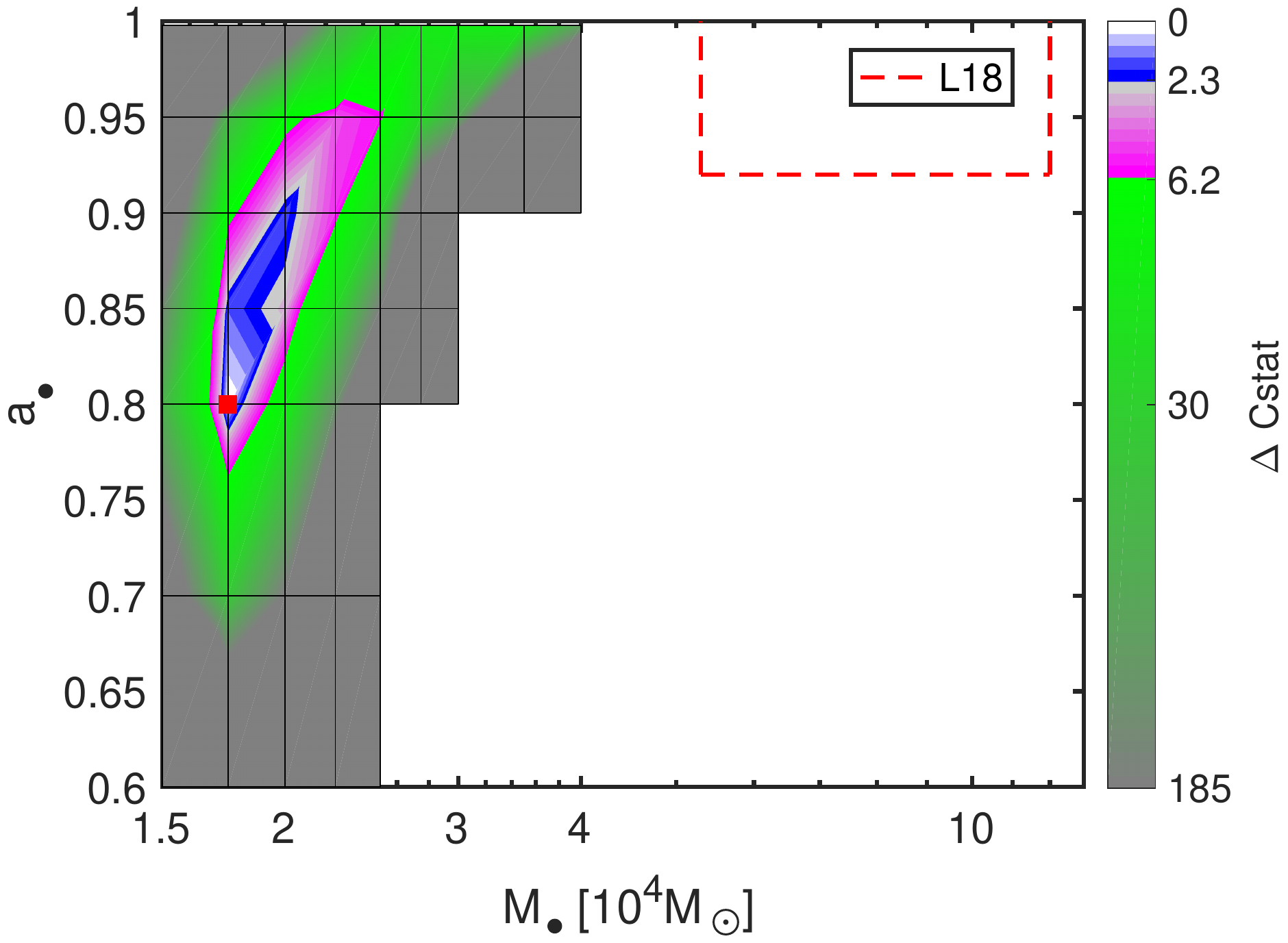}{0.5\textwidth}{Model 1: $f_{\rm c1}\in [2.0, 2.4]$ is allowed to float.}
\fig{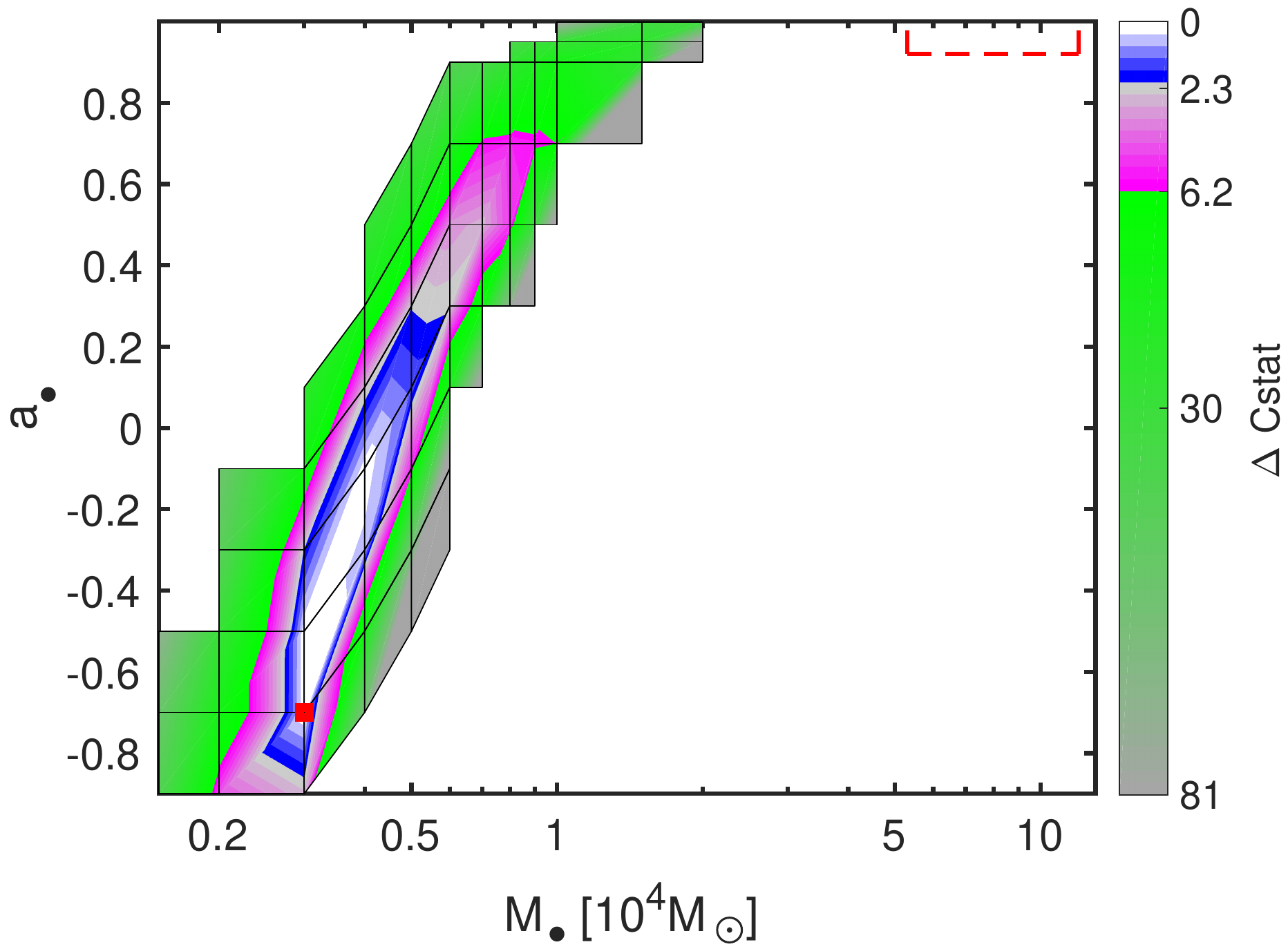}{0.5\textwidth}{Model 2: $\rm{z}$ is allowed to float. }
          }
\caption{Black hole mass $M_\bullet$ and spin $a_\bullet$ for our slim disk Model 1 and 2 (see text). We calculate the $\Delta Cstat$ across a model grid in the $(M_\bullet, a_\bullet)$ plane (grid points are indicated by vertices of the black lines) and then fill in the color contours by linear interpolation. The left panel shows one choice for modeling the spectral hardening factor $f_{\rm c}$, where it is fixed with a fiducial, theory-driven value \citep{DE18} for the sub-Eddington epochs for which it is calculated, but allowed to float for super-Eddington epochs (i.e., Epochs 1 and 2). The host redshift is taken to be that of the nearby, barred lenticular galaxy at $z=0.055$. The right panel shows the result of treating $z$ as a free parameter, but using the theoretical \citet{DE18} treatment of $f_{\rm c}$ at all epochs. The total $Cstat$ for the two models are 1124.36 and 1125.99, respectively. 
The left panel constrains $M_\bullet$ and $a_\bullet$ to $1.75^{+0.45}_{-0.05}\times 10^4 M_\odot$ and $0.8^{+0.12}_{-0.02}$, respectively, within 1$\sigma$.
The right panel constrains 
$M_\bullet$ and $a_\bullet$ to be $3.0^{+4.0}_{-0.5}\times 10^3 M_\odot$ and $-0.7^{+1.0}_{-0.1}$, respectively.
For the best fit $M_\bullet=3.0\times 10^3 M_\odot$ and $a_\bullet=-0.7$ in the right panel, we obtain a redshift of
$z=0.017 \pm 0.004$ ($1\sigma$ CL).
%
These two panels together show that regardless of uncertainties on $z$ and $f_{\rm c}$, this BH is an IMBH with mass less than $2.2\times 10^4 M_\odot$. The constraint on $M_\bullet$ is lower than that predicted by L18. Our spin constraint is the first of its kind.
}
\label{J2150}
\end{figure*}

For Model 1, we show the best fit results in
Fig.~\ref{J2150sp} and the left panel of Fig.~\ref{J2150} and the first section of Table~\ref{J2150p}. Here, $f_{\rm c1}$ is a free parameter with a flat prior between $2.0-2.4$. This prior is only applied to the first two epochs. For the last three epochs, $f_{\rm c}$ is calculated as in our fiducial method following the prescription provided in \citet{DE18}. The spectra, for each of the five epochs are well-fit with $Cstat/\nu$ $< 1.3$. The total $Cstat$ for the five epochs together is 1126.58, which is close to that expected ($Cstat=1124.1\pm1.4$; calculated following \citealt{Kaastra17}), indicating a good fit to the data. The mass accretion rate we derive to explain the Epoch 1 spectrum is highly super-Eddington even when using the high $f_{\rm c}$ value, while it is more mildly super-Eddington at Epoch 2. 

For Model 2, we list the best fit results in
the second section of Table~\ref{J2150p} and show the constraint on $M_\bullet$ and $a_\bullet$ in the right panel of Figure~\ref{J2150}. Here, the redshift is a free parameter (the redshift for our {\sc slimdisk} model and that of {\sc zphabs} are required to be the same). The best fit value of ($M_\bullet, a_\bullet)$ is $(3\times10^3M_\odot,-0.7$). The total $Cstat=1125.99$, similar to that of Model 1. The fit of Model 2 prefers a lower redshift, a lower $M_\bullet$, and a lower $a_\bullet$ than for Model 1. In other words, the observed spectrum can either be described by a softer intrinsic source at lower redshift or by a harder intrinsic source at higher redshift. A lower $M_\bullet$ implies a higher disk temperature, which is counteracted by the larger inner disk radius caused by the retrograde spin.

For comparison with the work of L18, we also fit the spectra with a simple {\sc diskbb} model. The fitted temperatures for the last four epochs are lower than in L18. This difference may come from the different statistics we employ, and/or from our different $N_{\rm{H}}$ treatment (they tie the $N_{\rm{H}}$ of four epochs together, while we let them float for each epoch). 
Our {\sc diskbb} fit yields a lower fitted $Cstat=1122.86$ with one more free parameter than the slim disk model fit. The better fit for the slim disk model can be explained by its improved fit to Epochs 1, 2, and 5. The best fit of the {\sc diskbb} model finds a larger absorption $N_{\rm {H}}=17.2\times10^{20}~ \rm{cm}^{-2}$ (10 times bigger than that of slim disk model) for Epoch 5. Even with stronger priors, e.g.,~requiring the same $M_\bullet$, the same $a_\bullet$ and the same $\theta$ for all epochs, the slim disk yields a better fit for the two early epochs, with $\Delta Cstat > 10$ lower than for the {\sc diskbb} model. The slim disk also finds a decreasing mass accretion rate, which is consistent with the expectation that the mass accretion rate in TDEs should decrease after rising to a peak.

\subsection{Event Parameters}


Fig.~\ref{J2150} shows the constraints on $M_\bullet$ and $a_\bullet$ for Models 1 and 2. Here, we assume $\Delta Cstat = 2.3$ and $\Delta Cstat = 6.2$ correspond to $1\sigma$ and $2\sigma$ confidence levels (CLs), respectively. For Model 1, we constrain $M_\bullet$ and $a_\bullet$ to $1.75^{+0.45}_{-0.05}\times 10^4$ $M_\odot$ and $0.8^{+0.12}_{-0.02}$, respectively. 
For Model 2, we  
constrain $M_\bullet$ and $a_\bullet$ to $3.0^{+4.0}_{-0.5} \times 10^3 M_\odot$ and $-0.7^{+1.0}_{-0.1}$, respectively. 
Model 2
rules out $z > 0.055$ at $>3\sigma$.
For the best fit $M_\bullet = 3\times10^3 M_\odot $ and $a_\bullet =-0.7$ from Model 2, the best fit $z$ is $0.017 \pm 0.004$ at $1\sigma$ CL.
The NASA Extragalactic Database (NED)\footnote{The NASA/IPAC Extragalactic Database (NED) is funded by the National Aeronautics and Space Administration and operated by the California Institute of Technology.} lists two faint galaxies projected within $\sim$8$^\prime$ of the TDE host and at photometrically-estimated
redshifts of $z=0.015$ and 0.019, implying offsets of $\sim$200 kpc.  Three other faint galaxies within $\sim$8$^\prime$ of the host lie at $z = 0.027$-0.028. It is possible that the star cluster hosting the TDE is associated with any of these galaxies, although none are as intrinsically bright or as near to the TDE host as the barred lenticular at $z=0.055$. 

For both Model 1 and 2, we find a lower $M_\bullet$ value than L18, who used a thin disk model. 
From the results in Fig.~\ref{J2150}, as well as the additional analyses in Appendix A and Fig.~\ref{fiduclai-fc}, we see that the L18 identification of the host as an IMBH is robust to both a range of choices regarding disk modeling (e.g.,~slim vs.~thin disk) and prescriptions for spectral hardening.  Although other TDE IMBH candidates have been identified in the past \citep[e.g.,][]{Maksym+14}, J2150 has the richest set of observational data and appears generally inconsistent with an SMBH origin.

If we do not assume anything about the redshift of the host, $a_\bullet$ is largely unconstrained: at a 2$\sigma$ CL, the Model 2 fit for $|a_\bullet|$ is consistent with both $0$ and $1$.  However, if we make the reasonable assumption that the host of J2150 is located at $z=0.055$, the redshift of the adjacent, barred lenticular galaxy, then $a_\bullet=0.8^{+0.12}_{-0.02}$.  IMBH spins have not been measured before, which makes J2150 a valuable object for testing different theories of IMBH formation and growth.  Different IMBH formation scenarios predict different IMBH spins at birth, some of which are compatible with the values inferred from Model 1 here.  For example, supermassive stars may form in runaway collisions of main sequence stars and then collapse due to subsequent general relativistic instability.  Simulations of this collapse process find high, but considerably sub-extremal spins; for example, \citet{ShibataShapiro02} find IMBH birth spins $a_\bullet \sim 0.75$, while \citet{Reisswig+13} find $a_\bullet \approx 0.9$ under substantially different collapse evolution.

Other IMBH formation/growth scenarios would be less compatible with the range of $a_\bullet$ we infer from Model 1.  For example, IMBHs may form with masses $M_\bullet \sim 10^2~M_\odot$ as the remnants of Pop~III stars in the early Universe \citep{MadauRees01, Greif+11} and then grow to larger sizes via gas accretion. If the incoming gas maintains a fixed orientation for timescales much longer than the Salpeter time, then the accreting IMBH seed will become nearly extremal in spin \citep{Thorne74}; conversely, if the angular momentum of the incoming gas randomizes its direction on timescales much shorter than the Salpeter time (the so-called ``chaotic accretion regime'' of \citealt{KingPringle06}), then the IMBH seed spins down to a spin in the range of $a_\bullet\sim 0.1-0.3$, depending on black hole mass and the details of disk realignment, with fluctuations between accretion episodes of $\Delta a_\bullet \pm 0.2$ \citep{King+08}.  In a cosmological context, growth of massive black holes through the chaotic accretion mode can produce even lower mean spin values \citep{BertiVolonteri08}.

This second latter option includes growth through stellar tidal disruptions. Recent cosmological simulations show that the high-redshift growth of IMBHs up to BH masses of $\sim 5 \times 10^5$ M$_\odot$ might indeed be dominated by TDEs (e.g., \citealt{pfister2020}). These outcomes are incompatible with the spin inference in Model 1, so if that measurement is correct, it rules out accretion- or TDE-driven growth of an IMBH seed with initial $M_\bullet \ll 10^4 M_\odot$, unless unusual conditions are met.  For example, one could fine-tune accretion-driven growth models to result in $a_\bullet \approx 0.8$ if the last $e$-fold of growth saw a large-scale accretion disk reverse its angular momentum over a timescale comparable to the Salpeter time.  Likewise, the estimated spin could be attained if the last e-fold of growth was driven by tidal disruption from a disk of stars with aspect ratio $\sim 0.1$. We note also that $a_\bullet \approx 0.8$ is compatible with the last e-fold of growth happening in a comparable-mass IMBH merger.

\begin{figure}[ht!]
\plotone{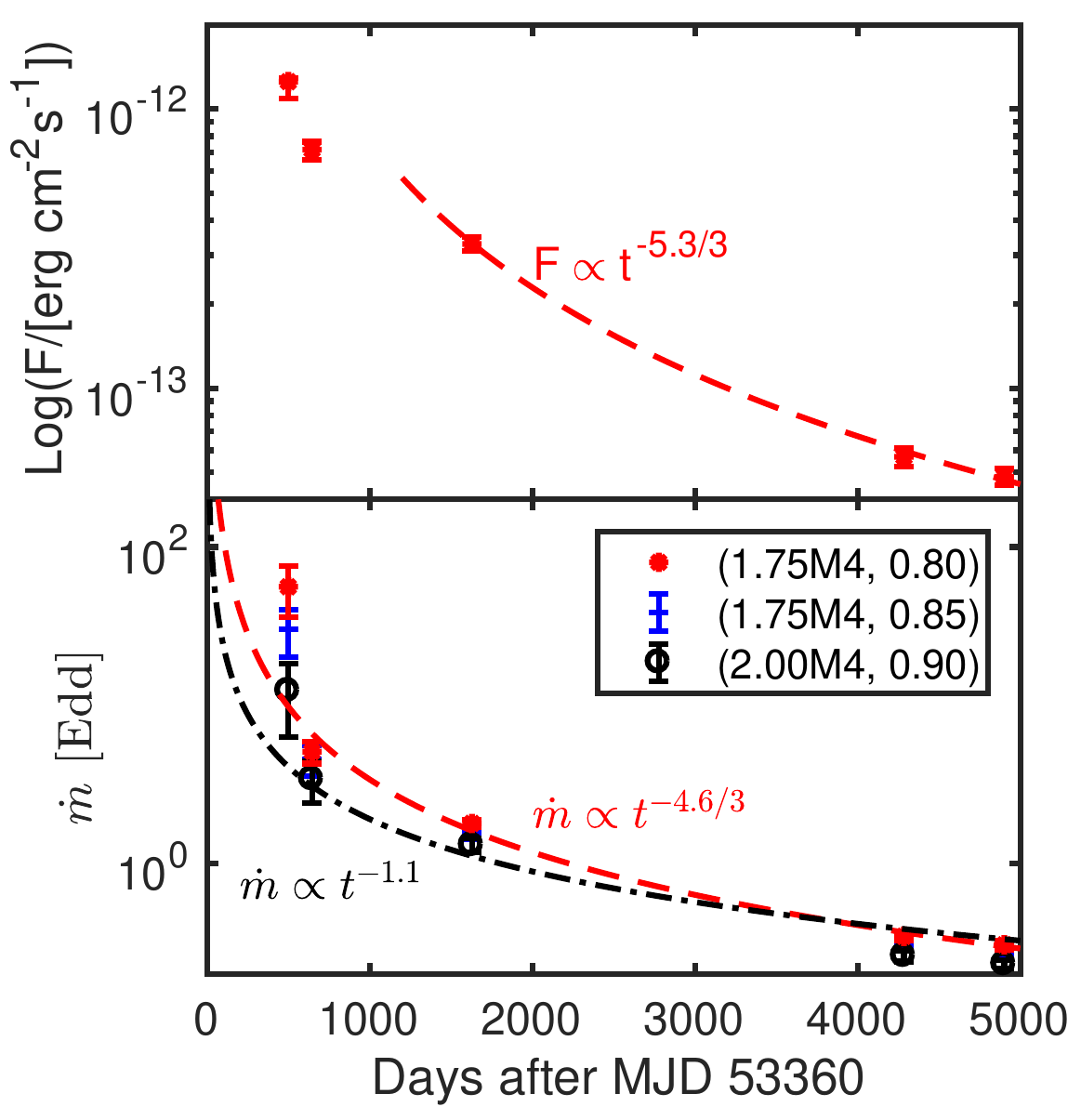}
\caption{The evolution of unabsorbed X-ray flux (top panel) and mass accretion rate (bottom panel). The error on the best fit flux is estimated by fixing the best fit inclination and absorption parameters, but varying only the accretion rate within the $1\sigma$ CL. Here, we set the peak date (arbitrarily) as 500 days prior to the first epoch of the X-ray observations. The red dashed lines represent the best fit power law to the last three epochs of the flux with $\chi^2/\nu=0.39/1$ ($ t^{-1.77\pm0.08}$) and to all five epochs of accretion rate with $\chi^2/\nu=10.78/3$. The disk accretion rate decays as $\propto t^{-1.53\pm0.06}$, steeper than that of $t^{-1.1}$ ($\chi^2/\nu=42.92/3$) in the TDE ASSASN-14li (W20) and marginally consistent with $t^{-5/3}$, indicating no significant 
circularization or viscous delay.  The late-time X-ray flux approximately traces the accretion rate.}  
\label{mfJ2150}
\end{figure}

Fig.~\ref{mfJ2150} shows the evolution of the unabsorbed X-ray flux and the disk accretion rate. Here and in the remainder of this section, we only consider the case of Model 1. The error on the measured flux is calculated by fixing the inclination and absorption, but adopting the lower and upper limit of $\dot m$ at $1\sigma$ CL. We plot the evolution of the X-ray flux for the best fit pair ($M_\bullet$, $a_\bullet$) and the evolution of accretion rate for the pairs ($M_\bullet$, $a_\bullet$) that fall within the $1\sigma$ contour in the left panel of Fig.~\ref{J2150}. The late time evolution of X-ray flux traces the decay of the accretion rate and is close to $t^{-5/3}$. This behavior is inconsistent with the prediction of an exponential decay from W20, because here we have an IMBH \citep{LodatoRossi11}. For IMBHs, the X-ray spectrum peaks at $\sim 0.8$ keV (see Fig.~\ref{J2150sp}), in contrast to the SMBH TDEs in W20, for which the 0.3 $-$ 7 keV band we observe is far down the Wien tail of the accretion disk. 

The accretion rate decays roughly as $t^{-1.53\pm0.06}$, which is also close to a $t^{-5/3}$ decay rate expected for late-time mass fallback. As a result, there should not be a significant viscous or circularization delay for this TDE during the epochs we observe. Indeed, the viscous timescale for an IMBH TDE disk is roughly $T_{\rm{vis}}=\alpha^{-1}\Omega^{-1}(2R_t)(H/2R_t)^{-2}\sim 0.1^{-1}\times 0.0036\times 0.3^{-2}~\rm{days}\sim 4.0 ~\rm{days}$, shorter than the gas fallback time scale $\sim 8.7 ~\rm{days}$ for $\beta=2$ ($\sim 7.8 ~\rm{days}$ for $\beta=1$).

The mass accreted by the IMBH during the five epochs is $(6.7\pm0.4)\times 10^{-3} ~M_\odot$, $(5.9\pm0.4)\times 10^{-3} ~M_\odot$ and $(4.8\pm0.4)\times 10^{-3} ~M_\odot$ for the three $(M_\bullet/M_\odot, a_\bullet)$ pairs $(1.75\times 10^4, 0.8)$, $(1.75\times 10^4, 0.85)$ and $(2 \times 10^4, 0.9)$, respectively. However, enhanced optical emission was seen from the TDE host roughly one year prior to the first X-ray observation (L18), suggesting that there may have been an earlier phase of significant accretion.  If we assume that the mass accretion peaked 500 days prior to the first epoch of the X-ray observations,
then the mass accreted from peak to Epoch 5 is $0.09\pm0.02 ~M_\odot$, $0.08\pm0.02 ~M_\odot$ and $0.05\pm0.02 ~M_\odot$, respectively (for the different mass-spin pairs given above). As a result, the flare is consistent with a full disruption if it maintained or exceeded its Epoch 1 luminosity for 1-2 years prior to the first \xmm\, observation, but would be consistent with a partial disruption (or the loss of most of the dynamically bound stellar debris in an outflow, as in \citealt{MetzgerStone16}) if the accretion rate rose quickly prior to the start of X-ray observations.

\subsection{Particle Physics Implications}

\begin{figure}[ht!]
\plotone{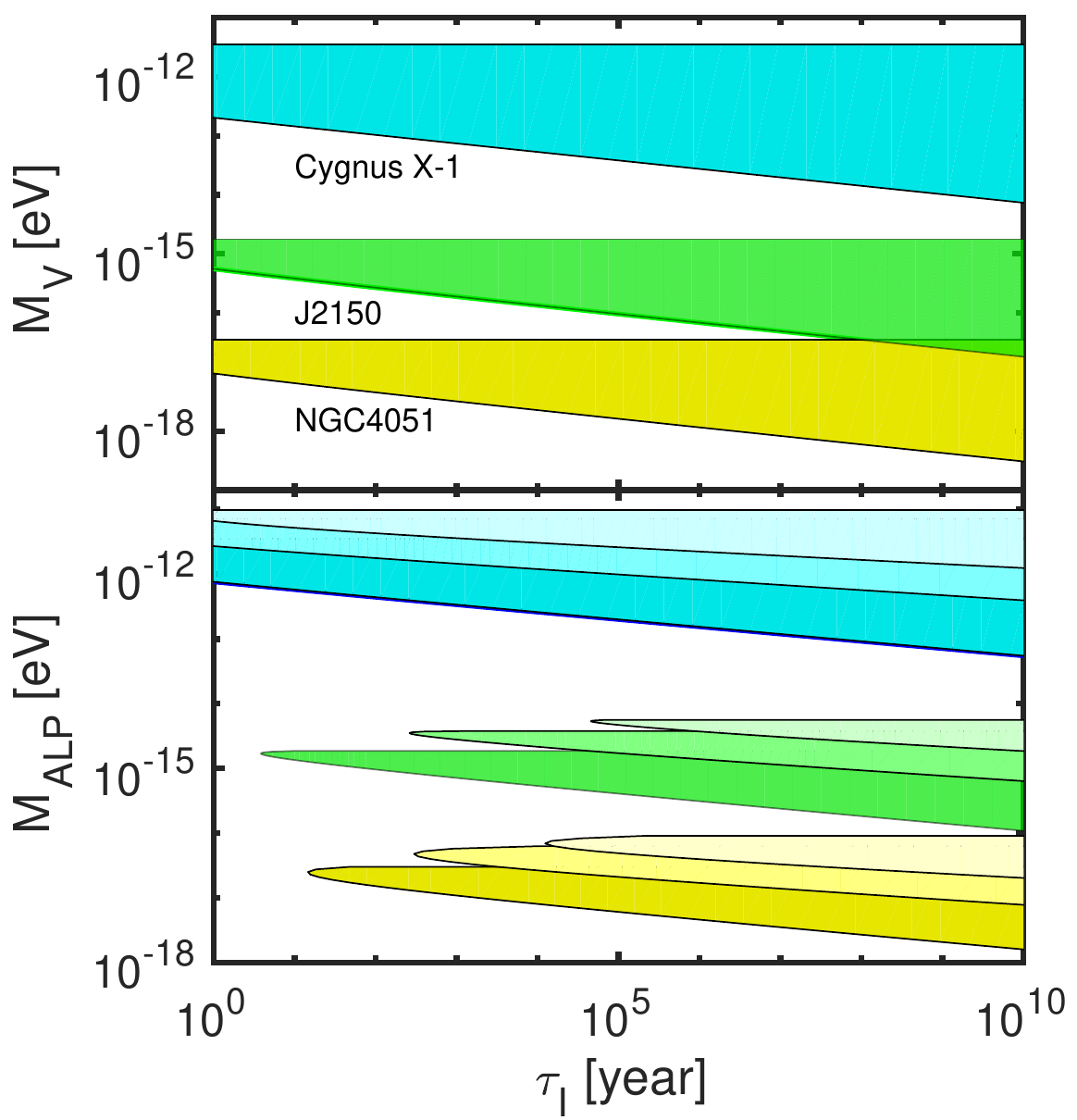}
\caption{Exclusion regions in the hypothetical masses of ultralight bosons, including both Proca vector bosons (mass $M_{\rm V}$) and scalar axion-like particles (mass $M_{\rm ALP}$), with our J2150 constraints (green). Both panels use the formalism of \citet{Cardoso+18} to show mass ranges of elementary particles excluded by astrophysical spin measurements of different black holes; in all cases the x-axis denotes the instability BH timescale, while the y-axis shows the particle mass. The upper panel shows the excluded mass of the Proca vector bosons, while the lower panel shows the excluded axion-like particle mass. The contours denote the excluded masses for a given $M_\bullet$, $a_\bullet$ and instability time $\tau_I$ (this can be interpreted as the astrophysical spin-up time for the most recent e-fold of growth of the BH in question; it cannot be less than the Salpeter time $\sim 4 \times 10^7$ yr for Eddington-limited accretion).  The green contours show the exclusion regions based on our mass and spin measurements for J2150.  The cyan and yellow contours denote the constraint of the ALP and Proca mass from the stellar BH system Cygnus~X-1 ($M_\bullet=14.80 M_\odot$, $a_\bullet=0.97$; \citealt{orosz2011}, \citealt{Parker+15}) and the SMBH system NGC~4051 ($M_\bullet=1.91 \times 10^6 M_\odot$, $a_\bullet=0.99$; \citealt{denney2009}, \citealt{patrick2012}). 
The dark, light and lighter contours in the lower panel denote the cases of low-order instability modes (modenumber $m=$1, 2 and 3, respectively). This figure shows that measurement of rapid spin in IMBHs, as we have performed for J2150, can exclude the existence of ultralight bosons at novel mass scales.}
\label{fig:ALP_V}
\end{figure}

If our Model 1 assumptions (source location at $z=0.055$; super-Eddington disk spectra can be modeled as a spectrally hardened multi-color disk blackbody) are correct, then the IMBH powering J2150 is rapidly spinning, with $0.78 < a_\bullet < 0.92$ at a 1$\sigma$ CL.  This represents the first spin measurement of an IMBH, and the high spin measured carries notable implications for particle physics.  In particular, rapidly rotating Kerr BHs are well known to exhibit a superradiant scattering instability in which spindown is triggered by interactions with ultralight bosons \citep{Bardeen+72, PressTeukolsky72}; the spin kinetic energy and angular momentum of the BH is converted into a bound cloud of elementary particles, saturating only after an order unity fraction of both have been transferred into the cloud \citep{EastPretorius17}.  This spindown instability is only efficient if bosons exist with mass $m$ such that $\frac{GM_\bullet m}{c\hbar} \sim 1$ (here $\hbar$ is the reduced Planck constant).  The timescale for linear growth of the instability grows exponentially if $\frac{GM_\bullet m}{c\hbar} \gg 1$ or $\ll 1$.

In practice, measurements of large $a_\bullet$ values in astrophysical BHs can be used to rule out roughly one order-of-magnitude in ultralight particle mass $m$ \citep[e.g.][]{Cardoso+18}. Figure 5 shows the excluded particle masses (colored regions) derived from the black hole masses and spins of Cygnus X-1 (blue), J2150 (green), and NGC 4051 (yellow). The black hole instability time $\tau_{\rm I}$ on the x-axis is calculated from Eq.~2.13 and Eq.~2.18 of \cite{Cardoso+18} for a given particle mass, $M_\bullet$, and $a_\bullet$. This is roughly equivalent to a “spindown time,” in that an isolated black hole will lose an order unity fraction of its spin over the timescale $\tau_{\rm I}$. When this spindown/instability timescale is much shorter than any plausible spinup timescale (which, for Eddington-limited accretion, would be of order the Salpeter time), an observed $(M_\bullet, a_\bullet)$ combination can be said to exclude a given boson mass. The stellar-mass BH system Cygnus X-1 excludes massive scalar fields (e.g., axion-like particles) and massive vector fields (e.g., dark photons) with $m\sim 10^{-12}$ eV, whereas the SMBH system NGC 4051 excludes those particles at $m\sim 10^{-17}$ eV. Spin measurements of larger SMBHs have been used in the past to exclude even smaller ultralight bosons \citep{Cardoso+18}. Critically, the mass and spin constraints that we derive here from J2150 (Model 1), an IMBH, exclude a new, intermediate particle mass range around $10^{-15}$ eV. For each astrophysical system, the range of excluded particle masses widens as instability timescale increases. 
%

To our knowledge, there have been no astrophysical spin measurements of IMBHs prior to our results; therefore, this is the first superradiance constraint on ultra-light boson masses. As such, it complements existing laboratory experiments, such as the CASPEr project \citep{Garcon+18}, which has already placed bounds on the existence of low-mass bosons in the $\sim 10^{-16}$-$10^{-13}$ eV range \citep{Garcon+19}. In contrast to laboratory experiments, however, superradiance constraints depend primarily on the mass of the boson and do not require any significant interactions with baryonic matter. Yet if the bosonic self-interaction is too strong, bound clouds formed through superradiance can self-annihilate \citep{YoshinoKodama12}, greatly reducing the degree of black hole spindown and potentially creating an observable gravitational wave signal \citep{Arvanitaki+15}. We do not consider the possibility of boson self-interaction and its effect on spindown rates here, except to note that if the boson self-interaction is too strong, superradiance constraints are weakened.  This scenario is treated in detail in \citep{Mathur+20}.


\section{Conclusions}
\label{sec:conclusions}

We have fit our general relativistic slim disk accretion model to the unusual TDE J2150.  Our approach is similar to our earlier work in W20, although here we improve our model to account for angular momentum lost by radiation.  We use a Kerr metric ray-tracing code to simultaneously fit 
five epochs of X-ray continuum spectra in J2150.
We explore fits with different priors to test the uncertainties 
in our model, including the assumed
disk outer and inner radius (Appendix E), spectral hardening parameterization (Appendix B), and TDE host redshift (Appendix C).  We find that:

\begin{enumerate}

\item Regardless of our choice of priors, we identify the central engine of the accretion disk to be an {\it intermediate-mass} black hole where
$M_\bullet$ is less than
2.2$\times 10^4 ~M_\odot $ at $1\sigma$. 

\item
If we assume the TDE host is associated with the adjacent, barred lenticular galaxy at $z=0.055$ (``Model 1''), we
achieve a good fit across 12 years of observations and two orders of magnitude in disk accretion rate. We
constrain the black hole mass $M_\bullet$ and spin $a_\bullet$ to be $1.75^{+0.45}_{-0.05}\times 10^4$ $M_\odot$ and $0.8^{+0.12}_{-0.02}$, respectively, at $1\sigma$. 
This high, but significantly sub-extremal, spin suggests that, if the IMBH has grown significantly since formation, it has acquired its last e-fold in mass in a way incompatible with both the ``standard'' and ``chaotic'' gas accretion limits, which predict spins that are too high and too low, respectively \citep{BertiVolonteri08}.
The spin
$a_\bullet$ depends sensitively on 
the unconfirmed
redshift of J2150's host system.
%
%
Measuring that redshift would 
eliminate a major systematic uncertainty on the spin measurement.

\item If our Model 1 is correct, we have discovered a rapidly spinning IMBH, the first measurement of its kind. The IMBH in J2150 would thus also represent the first ``superradiant scattering'' constraint on ultralight elementary bosons with masses $\sim 10^{-15}$ to $\sim 10^{-16}$ eV.  The existence of ultralight scalar (e.g.,~axion-like particles) or vector (e.g.,~dark photons) bosons in these mass ranges, respectively, can be ruled out due to the failure of the IMBH powering J2150 to spin down under the effects of superradiant scattering (although this conclusion is weakened for bosons with sufficiently strong self-interaction cross-sections).

\item
The flare is consistent with a full disruption if it maintained its Epoch 1 luminosity for 1-2 years prior to the first \xmm\, observation in May 2006, as is suggested by the 2005 identification of an optical outburst (L18).  If, however, the accretion rate rose quickly prior to the start of X-ray observations, then the accreted mass is very low ($\sim 10^{-3}$ to $10^{-2}M_\odot$) and requires either a partial disruption or the loss of most of the dynamically bound mass.

\end{enumerate}

In the near future, the X-ray satellites {\it SRG/eROSITA}, {\it Einstein Probe}, and possibly {\it Theseus} will together likely discover hundreds of new soft X-ray TDEs \citep{Khabibullin+14, Yuan+15, Jonker+19}. Targeted X-ray followup of TDEs found in optical surveys such as ZTF may find additional X-ray bright TDEs. Our analysis of J2150 demonstrates that if high-quality, multi-epoch X-ray spectra can be acquired for some of these TDEs, for instance through pointed \xmm, \chan, {\it SRG/eROSITA} or {\it Athena} observations, then it will be possible to map out the IMBH mass function and potentially constrain IMBH spins as well.  

\section*{Acknowledgements}

We thank Dacheng Lin for his help in providing X-ray spectra, offering useful advice on X-ray fitting in the early stages of this work and useful comments on the draft. We thank the anonymous referee for their helpful comments. We thank B.~Metzger, C.~Miller, A.~Loeb, and E. Kara for their guidance and suggestions. SW and AIZ thank Steward Observatory and the UA Department of Astronomy for post-doctoral support for SW.
NCS received support from the Israel Science Foundation (Individual Research Grant 2565/19), and thanks Lam Hui for informative conversations.
Our calculations were carried out at UA on the El Gato and Ocelote supercomputers, which are supported by the National Science Foundation under Grant No.~1228509. Our work here is partly based on observations obtained with XMM-Newton, an ESA science mission
with instruments and contributions directly funded by ESA Member States and NASA.
Our research has made use of data obtained from the Chandra Data Archive and software provided by the Chandra X-ray Center (CXC) in the application package CIAO.

\begin{appendix}

\begin{deluxetable*}{cccccccccccccc}
\tablecaption{Testing error estimation of parameters for C-statistic}
\tablewidth{0pt}
\tablehead{
\colhead{Spectrum }&&\colhead{counts }&&\colhead{$kT_{\chi^2}$ [keV] }&&\colhead{$1\sigma$ width$_{\chi^2}$ }&&\colhead{$kT_{Cstat}$ [keV]}&&\colhead{$1\sigma$ width$_{Cstat}$ }&&}
\startdata
$N_1=0.01$ && 452413 && $0.30045^{+0.00039}_{-0.00039}$ && 0.00078 && $0.30058^{+0.00039}_{-0.00039}$ &&0.00078\\
$N_2=0.001$ && 45067 && $0.30114^{+0.00126}_{-0.00124}$ && 0.00250 &&$0.30206^{+0.00125}_{-0.00125}$ &&0.00250 &&\\
$N_3=0.0001$ && 4411 && $0.28644^{+0.00365}_{-0.00359}$ && 0.00724 && $0.29650^{+0.00390}_{-0.00381}$ && 0.00771\\ 
Chandra 6791 && 4801 && $0.25021_{-0.00303}^{+0.00307}$ && 0.00610 && $0.26146^{+0.00330}_{-0.00318}$&& 0.00648 \\
Chandra 17862 && 152 && $0.12916_{-0.00704}^{+0.00758}$ && 0.01462 &&$0.13426_{-0.00618}^{+0.00665}$ && 0.01283 \\
\enddata
\label{stat}
\tablecomments{While the \chan\, data have a background component that we modeled separately (see text), the simulated data has no background component. Unlike the first three lines, the temperature of the blackbody fit to the \chan\, data decreases with time as the source evolves.}
\end{deluxetable*}

\section{Statistical analysis}
\label{cstatisitc}

\cite{Kaastra17} shows that the C-statistic can be used for 
assessing the goodness of fit of a spectral model 
and that it is
preferred for X-ray spectra, as the $\chi^2$ statistic gives biased results even for 20-30 counts per spectral bin. However, when estimating the error on the fit parameters for the C-statistic, the assumption is often made that the C-statistic converges to the $\chi^2$ statistic without confirming that there are sufficient number of counts per spectral bin and/or a sufficient number of spectral data bins to justify the Central Limit Theorem, which underlies the assumption that one can use the $\chi^2$-like statistical distribution of C-statistic values. In this work, we also assume that the C-statistic converges to the $\chi^2$ statistic, i.e., the $1\sigma$ errors on the fit parameters are determined by using $\Delta Cstat =1.0$ for single parameter and  $\Delta Cstat = 2.3$ for two parameters. Here, we check if the C-statistic indeed converges to the $\chi^2$ statistic when estimating the uncertainty regions on the best-fit parameters in this paper. 

We first generate three mock spectra using the \xmm ~pn response file and the {\sc XSPEC} {\sc bb} (blackbody) model, and then fit these spectra employing either the C-statistic or the $\chi^2$ statistic. The three spectra have the same temperature (kT=0.3 keV) and exposure time (1~ks), but a different normalization, e.g., 0.01, 0.001 and 0.0001 ($N_1$ through $N_3$ in Table~\ref{stat}). We only fit the spectra with the {\sc bb} model over the 0.3-2.0 keV band. The total number of X-ray photons is 452413, 45067, and 4411 in the 0.3-2.0 keV band for the three normalisations listed above, respectively. There are 342 spectral data bins for all three cases. $kT_{\chi^2}$ is determined by $\Delta \chi^2= 1$, while $kT_{Cstat}$ is determined by $\Delta Cstat= 1$ with the {\sc steppar} command in {\sc XSPEC}. 

We also extend our analysis to the two \chan ~spectra of J2150; see Table~\ref{stat} for the number of X-ray photons in those spectra. We fit the two spectra separately using the {\sc bb} model attenuated by absorption.  The model parameters used to describe the background are fixed at their best-fit value from the background-only spectral fits (see main text). We use the {\sc steppar} command to determine the $1\sigma$ CL on the best-fit value of kT at $\Delta \rm{statistic} = 1$ for both the C-statistic and $\chi^2$ statistic. Here, $N_{\rm H}$ is fixed at the best-fit value.

As we can see from Table~\ref{stat}, the value of the $1\sigma$ error is consistent when calculated using the $\chi^2$ and C-statistic. This result indicates that 
using $\Delta Cstat=1$ to
estimate the $1\sigma$ CL uncertainties on the best-fit value (single parameter) is justified and that the C-statistic converges to the $\chi^2$ statistic when estimating errors.
As the total number of detected X-ray photons decreases, using the $\chi^2$ statistic in the fit starts to bias the fit result, which is in line with the results of \cite{Kaastra17}.
    
\section{fiducial {\lowercase{$f_{\rm c}$}} treatment over different epochs}
\label{f-fc}
\begin{deluxetable*}{cccccccccccccc}
\tablecaption{Results from fitting five epochs of J2150 with our fiducial $f_{\rm c}$ slim disk model.}
\tablewidth{0pt}
\tablehead{
\colhead{ }&& \colhead{XMM 1}  && \colhead{Chandra 1} && \colhead{XMM 2}  && \colhead{Chandra 2} && \colhead{XMM 3} \\ 
\colhead{Date} && \colhead{2006-05-05}  && \colhead{2006-09-28} && \colhead{2009-06-07}  && \colhead{2016-09-14} && \colhead{2018-05-24}  
}
\startdata
Separate fitting \\
$N_{\rm H}$ $[10^{20}{\rm cm}^{-2}]$ && $2.4\pm2.4$ && $0.4\pm0.7$ && $6.2\pm2.3$ && $0^{+1.1}$ && $2.4\pm1.1$ \\
$\theta$ $[{^\circ}]$ && $5.0^{+17}_{-0}$ && $49.7\pm3.2$ && $5.0_{-0}^{+25}$ && $=\theta_3$ && $=\theta_3$ \\
$\dot m$ ${\rm [Edd]}$ && $8.9^{+18.8}_{-1.2}$ && $100_{-53}^{+0}$ && $14.9\pm8.0$ && $1.3\pm0.1$ && $1.2\pm0.7$ \\
$M_\bullet$ $[10^4{\rm M}_{\odot}]$ && $2.0$ && $1.5$ && $1.0$ && $=M_{\bullet,3}$ && $=M_{\bullet,3}$ \\
$a_\bullet$ && $0.9995$ && $0.9995$ && $-0.3$ && $=a_{\bullet, 3}$ && $=a_{\bullet, 3}$ \\
$Cstat/\nu$ && $297.78/270$ && $211.93/177$ && $273.64/283$ && $58.71/58$ && $268.79/240$ \\
\hline
Combining fitting \\
$N_{\rm H}$ $[10^{20}{\rm cm}^{-2}]$ && $2.7\pm0.6$ && $0.7\pm0.7$ && $1.0\pm0.4$ && $0^{+0.4}$ && $0^{+0.4}$ \\
$\theta$ $[{^\circ}]$ && $5.0^{+10}_{-0}$ && $=\theta_1$ && $=\theta_1$ && $=\theta_1$ && $=\theta_1$ \\
$\dot m$ ${\rm [Edd]}$ && $6.4\pm1.5$ && $2.0\pm0.2$ && $0.58\pm0.02$ && $0.135\pm0.007$ && $0.120\pm0.005$ \\ 
$M_\bullet$ $[10^4{\rm M}_\odot]$ && $2.25$ && $=M_{\bullet,1}$ && $=M_{\bullet,1}$ && $=M_{\bullet,1}$ && $=M_{\bullet,1}$ \\
$a_\bullet$  && $0.9995$ && $=a_{\bullet, 1}$ && $=a_{\bullet, 1}$ && $=a_{\bullet, 1}$ && $=a_{\bullet, 1}$ \\
$Cstat/\nu$ &&$300.01/271$ && $234.09/180$ && $274.91/283$ && $66.33/58$ &&$281.92/240$ \\
\enddata
\tablecomments{We adopt the same fit function as in Table~\ref{J2150p}, e.g., {\sc po + po + po + agauss + agauss + phabs(zphabs(slimdisk))}. Here, $f_{\rm c}$ is calculated by the fiducial method (see W20 and Appendix~\ref{f-fc}). The Milky Way $N_{\rm H}$ absorption is fixed at $2.6\times10^{20}$ ${\rm cm}^{-2}$ (L18). $N_{\rm H}$ shown in the table is thus associated with the TDE itself, its host star cluster, and possibly the nearby, barred lenticular galaxy at $z=0.055$. The error on the parameters are calculated with fixed $M_\bullet$ and $a_\bullet$.}
\label{J2150p2}
\end{deluxetable*}

\begin{figure*}[ht!]
\gridline{ \fig{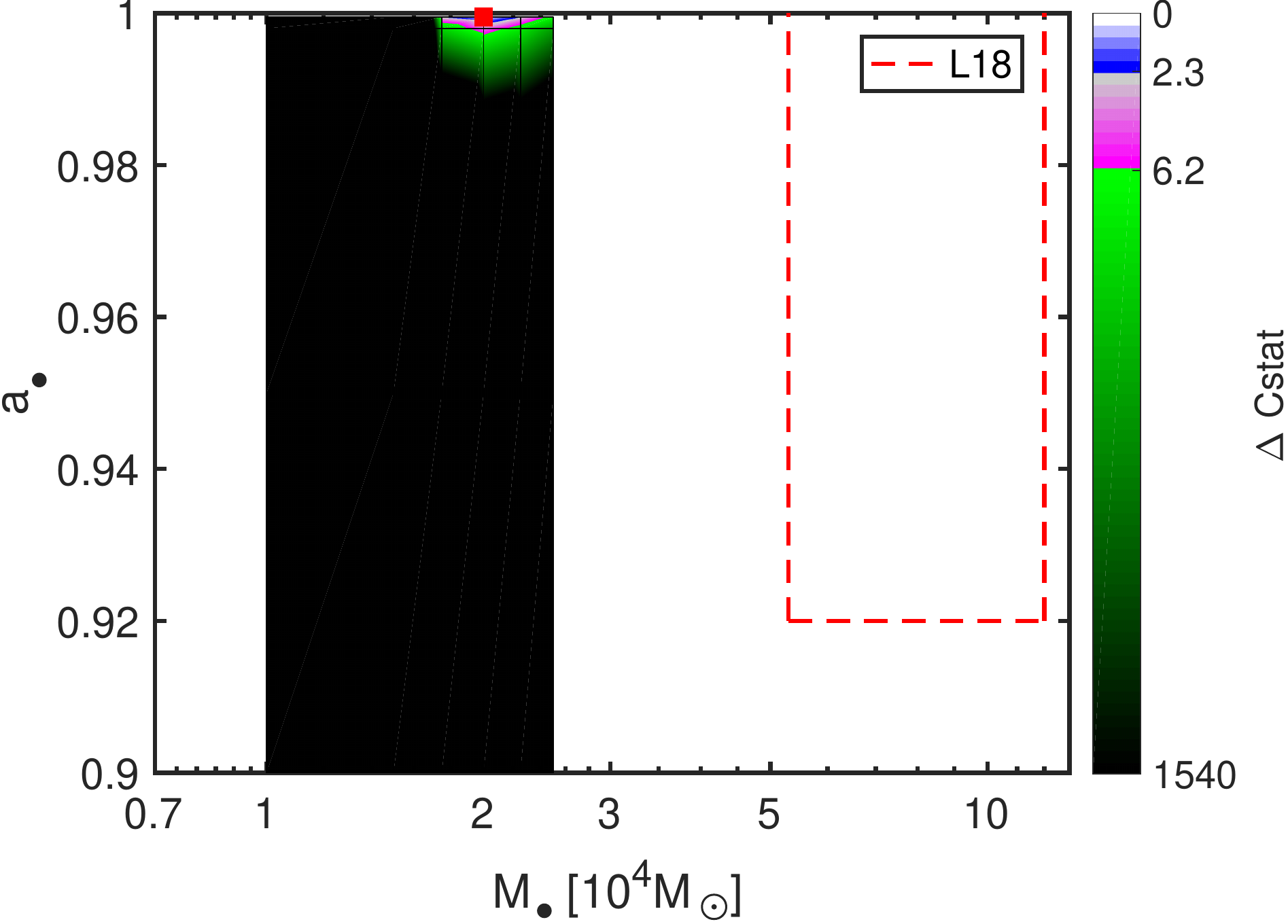}{0.5\textwidth}{Epoch 1}
\fig{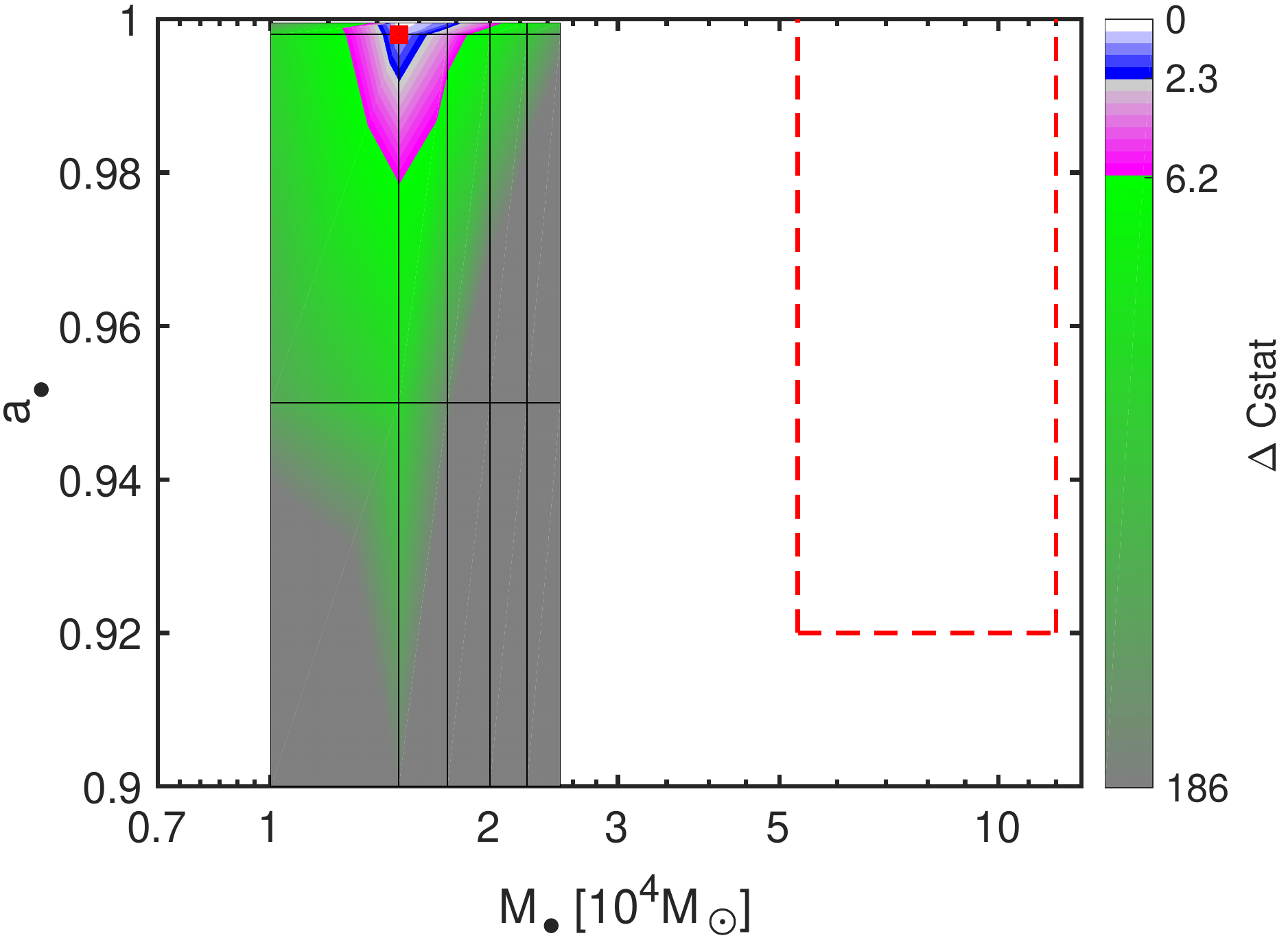}{0.497\textwidth}{Epoch 2 }
          }
\gridline{ \fig{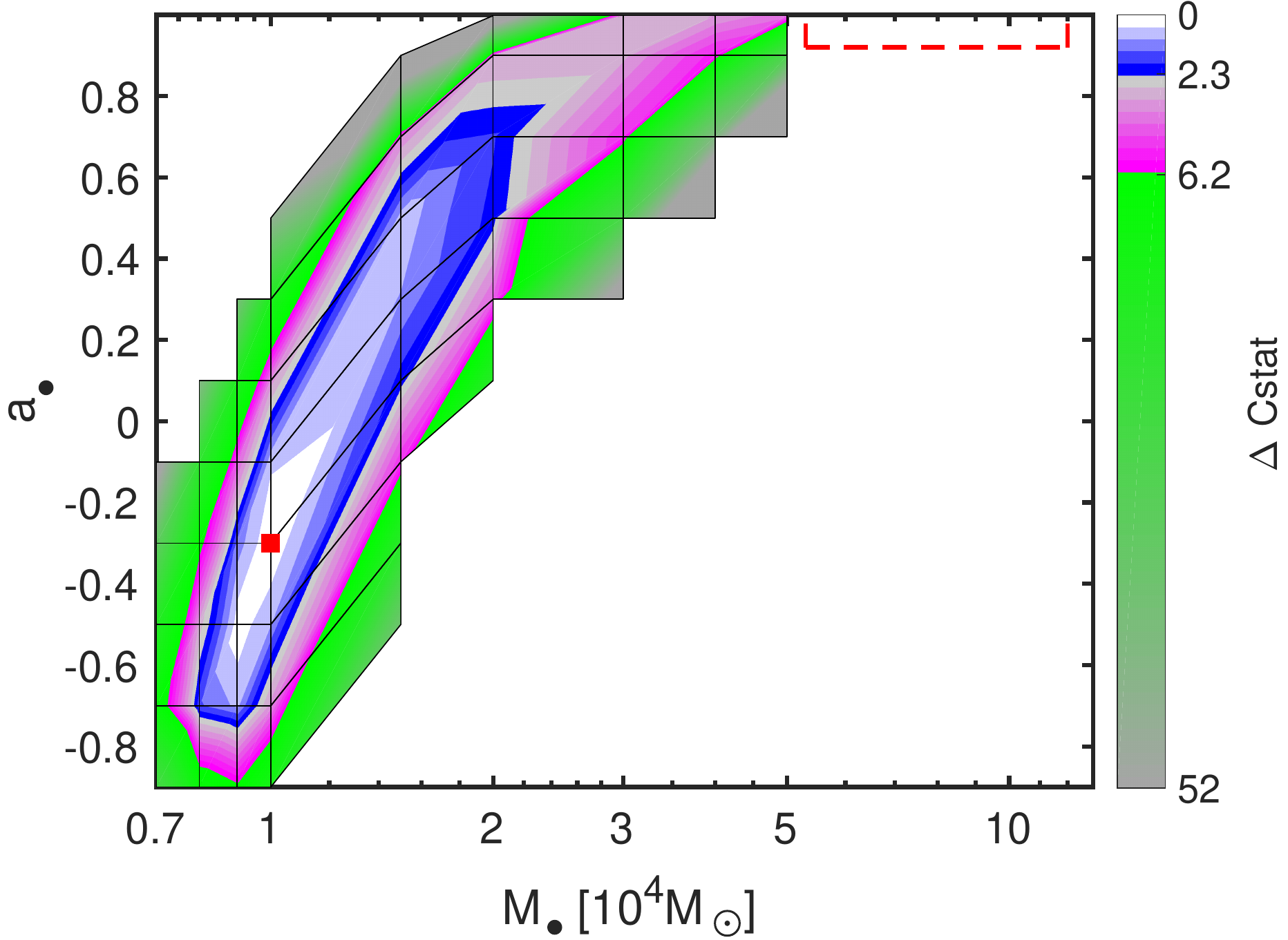}{0.495\textwidth}{Epoch 3-5}
\fig{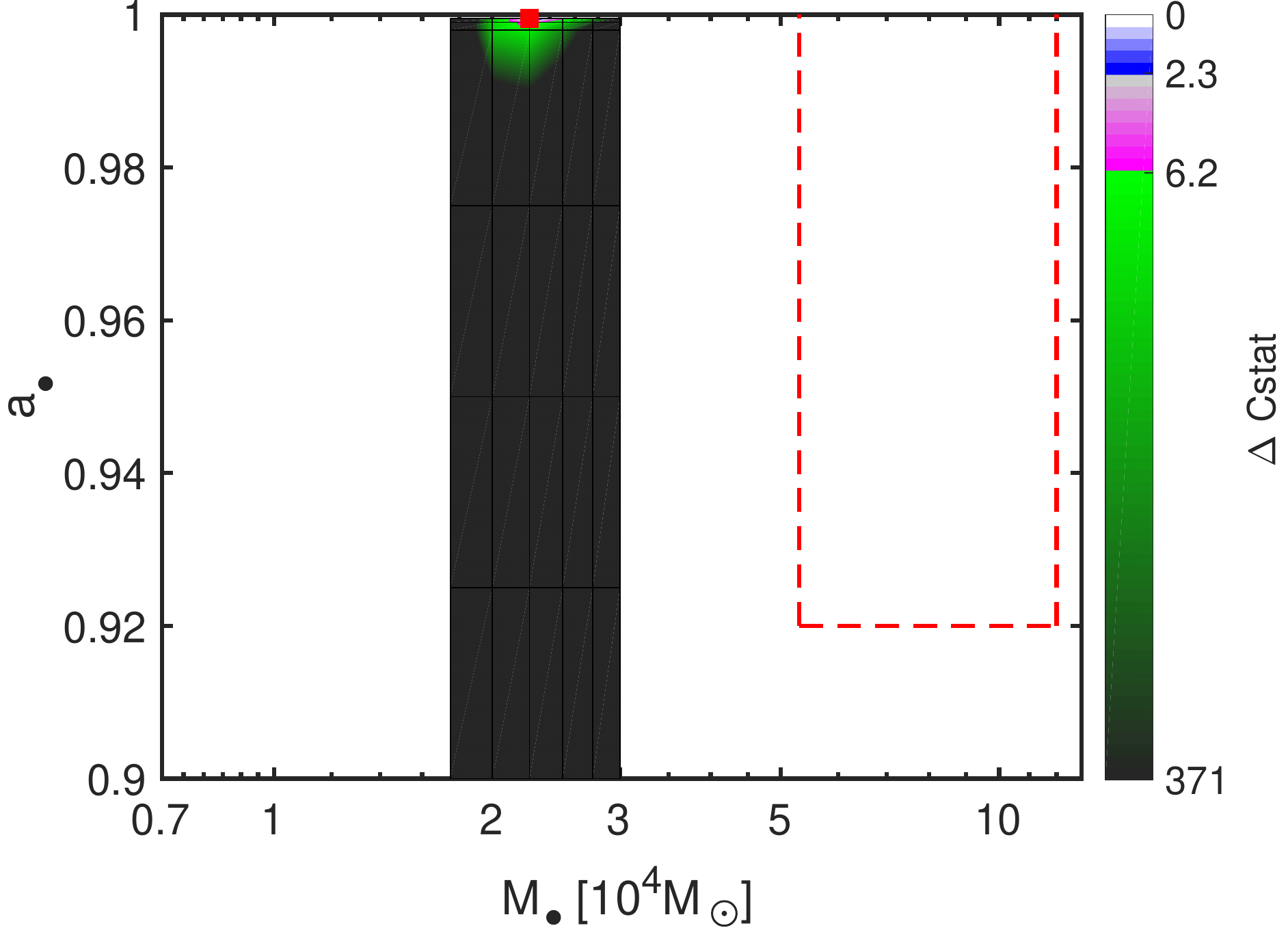}{0.5\textwidth}{Combine }
          }
\caption{For our fiducial $f_{\rm c}$, the effects of using different observed epochs to constrain $M_\bullet$ and $a_\bullet$ are shown. We calculate the $\Delta Cstat$ on a grid in the $(M_\bullet, a_\bullet)$ plane and then fill the intermediate parameter space by linear interpolation. The top left panel, top right panel, and the lower left panel show the results from Epoch 1, Epoch 2, and Epoch 3-5, while the lower right panel shows the result of the combining all five epochs. The best fit parameters of each panel are listed in Table~\ref{J2150p2}. The first two epochs produce constraints on $M_\bullet$ and $a_\bullet$ that are inconsistent with the later three epochs. The combined constraint on $M_\bullet$ and $a_\bullet$ is driven by the first two epochs. This inconsistency may arise from the underestimation of $f_{\rm c}$ for the early two super-Eddington spectra, Epochs 1 and 2.  
}
\label{fiduclai-fc}
\end{figure*}

Early studies \citep{ST93,ST95} showed that the local X-ray flux at each annulus will be higher than the corresponding blackbody flux, due to electron scattering and the temperature gradient in the atmosphere. The local X-ray emission can be approximated by a color-corrected blackbody \citep{ST95},
\begin{equation}
\label{Iv}
 I(\nu)=\frac{2h\nu^3c^{-2}f_{\rm c}^{-4}}{\exp(h\nu/k_{\rm B} f_{\rm c}T)-1}.
\end{equation}
Here, $h$ and $k_{\rm B}$ are the Planck constant and the Boltzmann constant, respectively; $f_{\rm c}$ is the spectral hardening factor. For a sub-Eddington disk, $f_{\rm c}$ is about 1.7 and insensitive to disk parameters \citep{ST95}. Later studies showed that $f_{\rm c}$ may increase with accretion rate \citep{GD04,DBHT05}. \citet{DE18} estimated $f_{\rm c}$ for a non-spinning SMBH,
\begin{eqnarray}
\nonumber 
f_{\rm c}=1.74+&&1.06(\log_{10} T-7)-0.14[\log_{10} Q-7] \\ 
&&-0.07\{ log_{10}[\Sigma/2]-5 \}. 
\label{fc}
\end{eqnarray}
Here, $Q$ and $\Sigma$ are the strength of vertical gravity and surface density at each annulus of the disk, respectively. This  $f_{\rm c}$ estimate holds for accretion rates between 0.01 to 1 Eddington units. In the super-Eddington regime, $f_{\rm c}$ would not increase to infinity as accretion increases and would instead saturate \citep{Davis2006} at about 2.4 for a SMBH accretion disk. In this paper, we take $f_{\rm c}$ from Eq.~\ref{fc} as our fiducial $f_{\rm c}$ treatment. As this fiducial $f_{\rm c}$ may not work well for a highly super-Eddington accretion disk, we parameterize $f_{\rm c}$ and set a flat prior of (2.0, 2.4) in that case.

In this section, we examine further the results of fitting the spectra with our fiducial $f_{\rm c}$ treatment and the problems that arise from this assumption. We first fit the spectral epochs separately, constraining the corresponding $M_\bullet$ and $a_\bullet$. We divide the five spectra into three groups, with Epoch 1 and Epoch 2 as two separate groups and Epoch 3-5 as the third group.  We break up our analysis in this manner, because it is unclear whether the fiducial $f_{\rm c}$ prescription of \citet{DE18} can be successfully extrapolated beyond the sub-Eddington regime in which it was derived, and applied to strongly super-Eddington accretion rates such as those in Epochs 1 and 2.  We also perform a simultaneous fit to all five epochs with the fiducial \citet{DE18} prescriptions.

Figure \ref{fiduclai-fc} shows the fitting results, and the corresponding best fit parameters are listed in Table~\ref{J2150p2}. Epoch 1 and Epoch 2 yield very narrow contours in the ($M_\bullet$, $a_\bullet$) plane. The best-fit $a_\bullet$ is pushed to an extremely high value beyond theoretically predicted saturation spins \citep[e.g.][]{Thorne74}, $\approx 0.9995$. Epoch 3-5 yields contours in ($M_\bullet$, $a_\bullet$) similar to ASSASN-14li (W20), but with even less constraint on $a_\bullet$. Our separate analysis of these three groups with the fiducial \citet{DE18} $f_{\rm c}$ prescription have consistent constraints on $M_\bullet$, but not on $a_\bullet$. The $1\sigma$ contours of Epoch 1 and Epoch 2 are at least $3\sigma$ away from those of Epoch 3-5. 

The combined Epoch 1-5 ($M_\bullet$, $a_\bullet$) constraints are driven by the first two epochs. $Cstat$ increases very quickly as spin decreases, indicating that the disk of a low spin BH is not bright enough to fit the early observations, which may be caused by an underestimated $f_{\rm c}$ in the fiducial model. From the separate fits, we see that both two early epochs are indeed in a highly super-Eddington phase (for $a_\bullet=0.9995$, the radiation efficiency is $\eta\approx 0.36$).  In W20, we showed that X-ray flux would be nearly constant in the highly super-Eddington regime. As a result, the X-ray luminosity of the disk is insensitive to accretion rate, but sensitive to the choice of $f_{\rm c}$. The 
unusual behavior of the $M_\bullet$ and $a_\bullet$ contours in the combined fitting may arise from the underestimation of $f_{\rm c}$ for a low spin disk. Motivated by (1) the lack of theoretical calculations for super-Eddington $f_{\rm c}$ values and (2) the incompatibility between the Epoch 1/Epoch 2 and Epoch 3-5 $a_\bullet$ constraints under \citet{DE18} $f_{\rm c}$ prescriptions, we allow $f_{\rm c}$ to float as a free fit parameter for the first two epochs. 

\section{effects of redshift uncertainty}
\label{app:simulation}
\begin{figure}[ht!]
\plotone{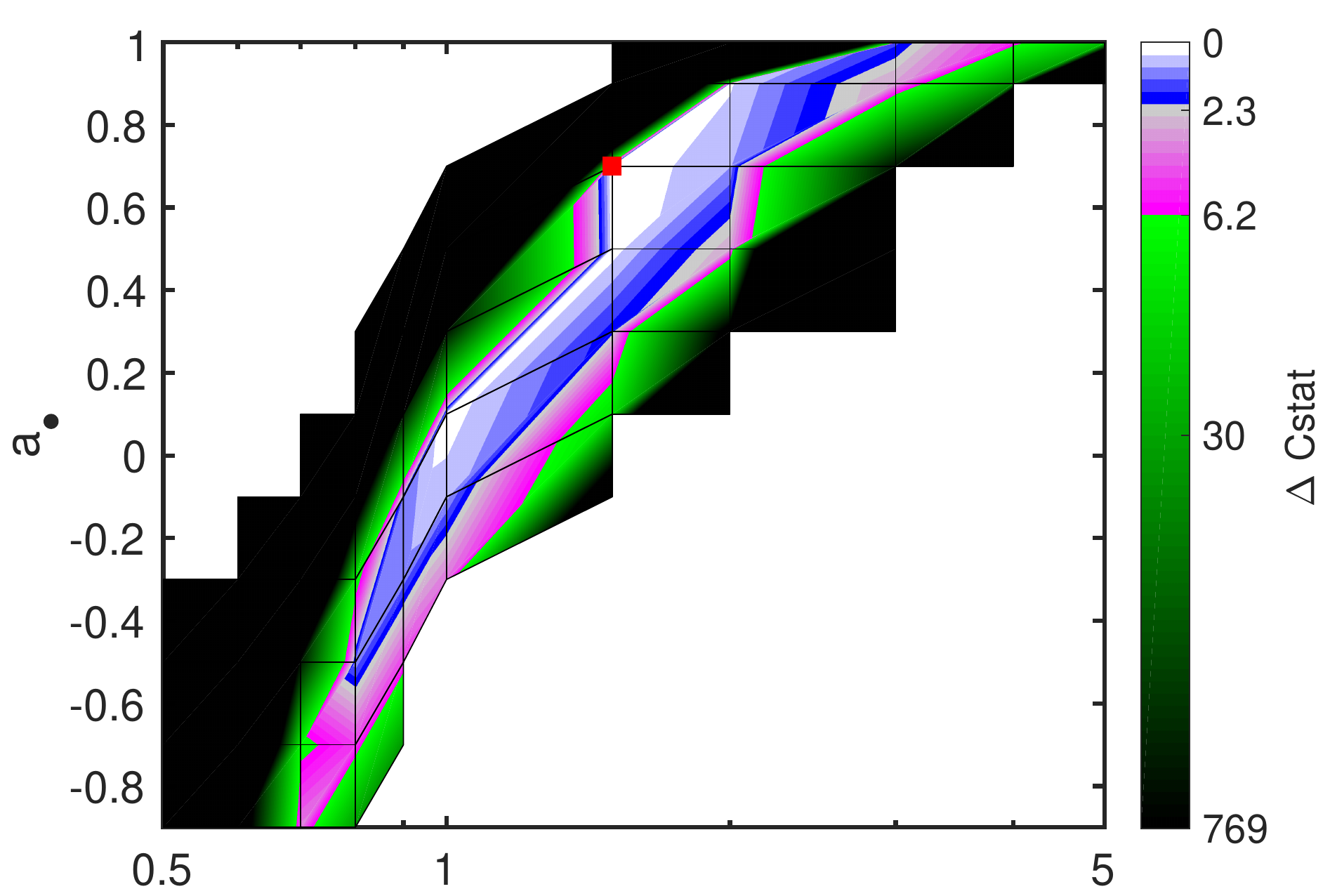}
\plotone{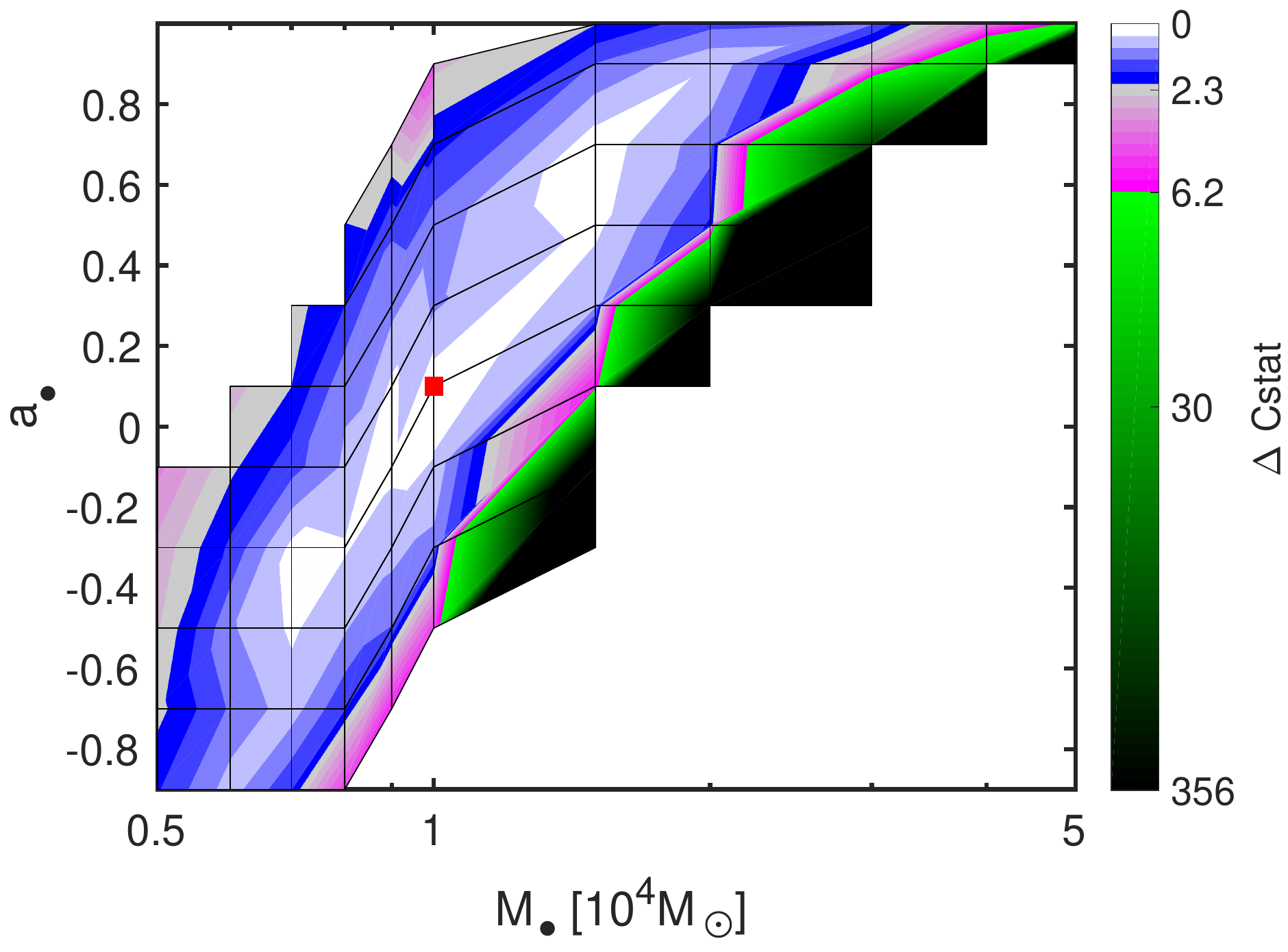}
\caption{Constraints on $M_\bullet$ and $a_\bullet$ from varying the TDE redshift assumption. We generate two mock spectra with the pn respond file with slim disk model. The value of parameters are $M_\bullet=2 \times 10^4 ~M_\odot$, $a_\bullet=0.9$, $N_{\rm{H}}=3.0\times10^{20} ~\rm{cm}^{-1}$, $\theta=10^\circ$, $z=0.055$, $\dot m_1 = 1$, and $\dot m_2 = 0.3$. The upper panel fits the spectra with $z=0.055$, while the lower panel allows $z$ to float. The best $Cstats$ for these two models are 819.08 and 818.94, respectively, slightly preferring a smaller $M_\bullet$, a lower $a_\bullet$, and a lower $z$. 
}
\label{simulation}
\end{figure}

In the section, we explore the effect of the unknown TDE redshift on our ($M_\bullet$, $a_\bullet$) constraints. We first generate two mock spectra with the XMM pn response file using our slim disk Model 1, and then refit the spectra using different settings of $z$. We generate the spectra with parameters $M_\bullet=2 \times 10^4 M_\odot$, $a_\bullet=0.9$, $N_{\rm{H}}=3.0\times10^{20} \rm{cm}^{-2}$, $\theta=10^\circ$, $z=0.055$, $\dot m_1 = 1$ Edd, and $\dot m_2 = 0.3$ Edd. The exposure time for both spectra is 200,000 seconds, and the number of counts in the spectra are 39,927 and 11,389 counts in band 0.3 -- 7.0 keV, respectively.

Fig.~\ref{simulation} shows the results of these fits. In the upper panel, we fit the spectra by fixing $z=0.055$. Within the $1\sigma$ contour is the pair value ($M_\bullet=2 \times 10^4 ~M_\odot$, $a_\bullet=0.9$), which we have used to generate the spectra. From the contours, we see that $M_\bullet$ is degenerate with $a_\bullet$ (W20). This degeneracy arises from the fact that either smaller $M_\bullet$ or higher $a_\bullet$ can produce a hotter disk.
For the lower panel, we use the same fit function as the upper panel, although in addition we treat the redshift $z$ as a free floating parameter in the fit. The best fit $Cstat$ is 818.94, very close to that of 819.08 from the upper panel fit. Therefore, allowing $z$ to float does not improve the fit significantly. However, the $M_\bullet$ and $a_\bullet$ contours become bigger, e.g., more models with smaller $M_\bullet$ and lower $a_\bullet$ can describe the spectra well. We also find that $z$ can be smaller than 0.055; e.g., the best fit is $z=0.033$ for $M_\bullet=5 \times 10^3 M_\odot$ and $a_\bullet=-0.9$. A lower $z$ is always associated with a lower $M_\bullet$, as a lower $z$ makes the expected spectra brighter by reducing the luminosity distance and redshifting the spectra less, while a lower $M_\bullet$ requires the disk to be hotter, increasing the flux as well as blueshifting the spectra. 
Therefore, a smaller $M_\bullet$ and a lower $z$ may arise from the degeneracy of the $z$, $M_\bullet$ and $a_\bullet$ parameters.

\section{Stationary slim disk model}
\label{app:slimdisk}
We adopt the same procedure as in W20 to reduce the relativistic slim disk equations. The equations of state, vertical hydrostatic equilibrium \citep{Abramowicz+1997}, and mass conservation are the same as that in W20. For brevity, the aforementioned equations are not detailed in this appendix, and we refer the reader to W20 for more detail. Here, we only write the three equations:
(1) angular momentum conservation equation \citep{Abramowicz+1996},
\begin{equation}
\frac{\dot{M}}{2\pi}\frac{d {\cal L}}{{\rm d}r}-\frac{{\rm d}}{{\rm d}r}\left(\frac{\alpha A^{1/2}\Delta^{1/2}\gamma P}{r}\right)=Q^{\rm rad}{\cal L};
\label{L}
\end{equation}
(2) radial momentum conservation equation,
\begin{equation}
\frac{V^2}{1-V^2}\frac{{\rm d}\ln V}{{\rm d} r}=\frac{{\cal A}}{r}-\frac{P}{\Sigma}\frac{{\rm d}\ln P}{{\rm d} r};
\label{dvdr}
\end{equation}
(3) energy conservation equation,
\begin{equation}
Q^{\rm adv}=-\alpha P\frac{A\gamma^2}{r^3}\frac{{\rm d}\Omega}{{\rm d}r }- \frac{64\sigma T_{\rm c}^4}{3\Sigma\kappa}. 
\label{Qadv}
\end{equation}
All the parameters are defined the same as in W20. These equations can be simplified to:
\begin{eqnarray} 
\label{2D1}
&&\frac{{\rm d}\ln V}{{\rm d} r}=\frac{N}{D}=\frac{A_3B_2-B_3A_2}{A_1B_2-A_2B_1}, \\
\label{2D2}
&&\frac{{\rm d}\ln T_{\rm c}}{{\rm d}r}=\frac{B_3}{B_2}-\frac{B_1N}{B_2D},\\
&&c_3\frac{{\rm d}\ln \cal L}{{\rm d} r}=c_4-c_1\frac{{\rm d}\ln V}{{\rm d} r}-c_2\frac{{\rm d} \ln T_{\rm c}}{{\rm d} r},
\end{eqnarray}
where
\begin{eqnarray}
\nonumber
&&A_1=a_1c_3-a_3c_1, A_2=a_2c_3-a_3c_2, A_3=a_4c_3-a_3c_4,\\ \nonumber
&&B_1=b_1c_3-b_3c_1, B_2=b_2c_3-b_3c_2, B_3=b_4c_3-b_3c_4.
\end{eqnarray}
Here $a_i$, $b_i$ and $c_i$ (i=1,2,3,4) are function of $T_{\rm c}$, $V$, $\cal{L}$ and r. They can be written as:
\begin{eqnarray}
\nonumber
&&a_1=P_1+\frac{V^2 \Sigma}{(1-V^2)P}, ~~a_2=P_2, \\ \nonumber
&&a_3=P_3, ~~a_4=\frac{{\cal A} \Sigma}{rP}-P_4,\\ \nonumber
&&b_1=\frac{4-3\beta_{\rm{p}}}{1-V^2}-\frac{2\pi\alpha A\lambda^2\Sigma}{\dot Mr^2}O_1,\\ \nonumber
&&b_2=12-10.5\beta_{\rm{p}}, ~~b_3=-\frac{2\pi\alpha A\lambda^2\Sigma}{\dot Mr^2}O_2,\\
\nonumber
&&b_4=\frac{2\pi\alpha A\lambda^2\Sigma}{\dot Mr^2}O_3+\frac{2\pi rQ^{rad}\Sigma}{\dot M P}-\frac{4-3\beta_{\rm{p}}}{2}\frac{{\rm d} \ln\Delta}{{\rm d}r}, \\
\nonumber
&&c_1=-\lambda_1-P_1, ~~c_2=-P_2, \\ \nonumber
&&c_3=\frac{\dot Mr\cal L}{2\pi\alpha \sqrt{A\Delta}\lambda P}-P_3-\lambda_2,\\ \nonumber
&&c_4=\frac{Q^{rad}r^2\cal L }{\alpha \sqrt{A\Delta}\lambda P}+P_4+\lambda_3+\frac{{\rm d} \ln(\sqrt{A\Delta}/r)}{{\rm d}r},\\
\nonumber
&&\lambda_1=\frac{V^2}{(1-V^2)^2\lambda^2},~\lambda_2=\frac{{\cal L}^2r^2}{A\lambda^2},\\\nonumber
&&\lambda_3=\lambda_2\left(\frac{1}{r}-\frac{\rm{d}\ln A}{2\rm{d}r}\right),\\
\nonumber
&&O_1=-(\Omega-\omega)\lambda_1,~O_2=(\Omega-\omega)(1-\lambda_2),\\\nonumber
&&O_3=\frac{{\rm d}\omega}{{\rm d}r}+(\Omega-\omega)(\frac{3}{r}+0.5\frac{\rm{d}\ln \Delta}{{\rm d}r}-\frac{3}{2}\frac{\rm{d}\ln A}{{\rm d}r}-\lambda_3), \\
\nonumber
&&P_1=-\frac{3\beta_{\rm p}-1}{1+\beta_{\rm p}}\frac{1}{1-V^2}-\frac{\beta_{\rm p}-1}{1+\beta_{\rm p}}\frac{a^2u_t(u_t+\omega{\cal L})\lambda_1}{r^4\cal G },\\ \nonumber
&&P_2=\frac{8-6\beta_{\rm p}}{1+\beta_{\rm p}}, \\
\nonumber
&&P_3=\frac{\beta_{\rm p}-1}{1+\beta_{\rm p}}\left(2-a^2\frac{u_t^2-1 -u_t((u_t+\omega{\cal L})\lambda_2-\omega{\cal L})}{r^4\cal G}\right),\\
\nonumber
&&P_4=-\frac{1}{2}\frac{3\beta_{\rm p}-1}{1+\beta_{\rm p}}\frac{{\rm d}\ln \Delta}{{\rm d}r}-\frac{\beta_{\rm p}-1}{1+\beta_{\rm p}}\frac{4}{r} -\frac{\beta_{\rm p}-1}{1+\beta_{\rm p}}\frac{a^2u_t}{r^4\cal G}\times\\ \nonumber
&&\left((u_t+\omega{\cal L})(\frac{{\rm d}\ln\Delta}{2{\rm d}r}-\frac{{\rm d}\ln A}{2{\rm d}r}+\frac{1+r\lambda_3}{r})-\frac{{\rm d}\omega}{{\rm d}r}{\cal L}\right).
\end{eqnarray}

We estimate the initial conditions by assuming the Novikov-Thorne disk ($\Omega=\Omega_k^+$, $Q^{\rm adv}=0$ and $Q^{rad}{\cal L}=0$). As a result, angular momentum conservation and energy conservation equations can be rewritten as, 
\begin{eqnarray}
&&\frac{\dot{M}}{2\pi}({\cal L}-{\cal L}_{\rm in})=\frac{A^{1/2}\Delta^{1/2}\gamma}{r}\alpha P,\\
&&\alpha P\frac{A\gamma^2}{r^3}\frac{{\rm d}\Omega}{{\rm d}r }=\alpha P\frac{A\gamma^2}{r^3}\frac{{\rm d}\Omega_k^+}{{\rm d}r }=- \frac{64\sigma T_{\rm c}^4}{3\Sigma\kappa}.
\end{eqnarray}

Combined with other equations, one can solve $V$, $T_{\rm c}$ and $\cal {L}$ for a given r. Here, ${\cal L}_{in}$ is the integration constant, which denotes the angular momentum component at the disk inner edge. The free parameters are $M$, $a$, $\dot M$ and $\alpha$. ${\cal L}_{in}$ is the eigenvalue of the problem, which must be chosen properly to ensure that $N=0$ and $D=0$ at the sonic point. We use the shooting technique to narrow ${\cal L}_{in}$ (W20). The ${\cal L}_{in}$ estimation is updated iteratively, until $\Delta {\cal L}_{in}/{\cal L}_{in}$ is less than $10^{-6}$. We iteratively integrate the equations to radius near the sonic point with the latest ${\cal L}_{in}$ estimate, then take a large step ahead and continue to solve the equations to near-horizon distances. Our solutions are insensitive to the initial conditions.  

\begin{figure}[ht!]
\plotone{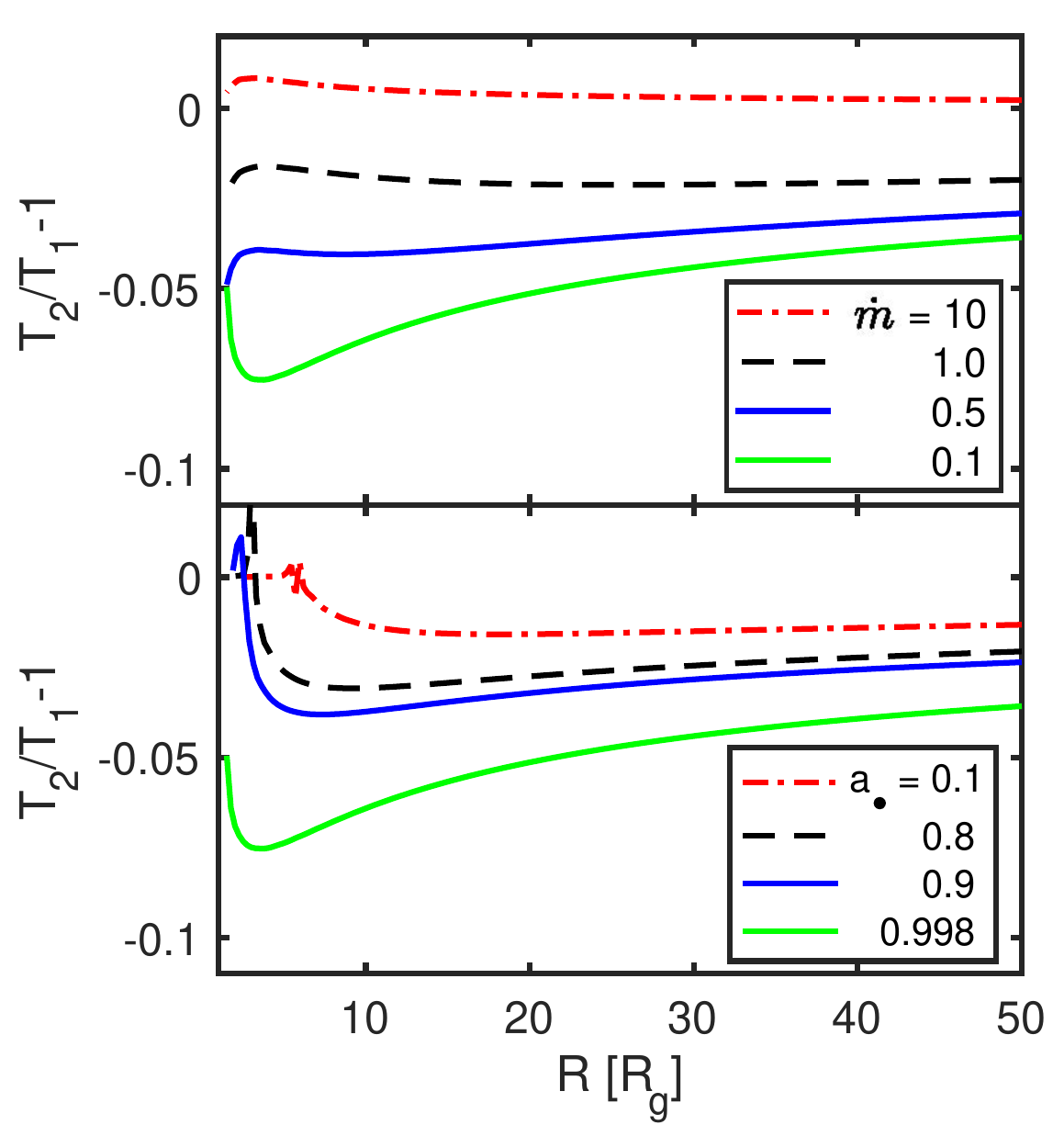}
\caption{The comparison of effective temperature $T(R)$ between disk models that account for angular momentum lost in radiation ($T_2$) and those that do not ($T_1$). In both panels, we plot the temperature difference against dimensionless radius $R$.  The figure show that the angular momentum loss effect is strongest for high spin and low accretion rate disks.  For $a_\bullet<0.8$, the effect is weak, and the error is $< 5\%$.
}
\label{AM}
\end{figure}

Figure \ref{AM} shows the comparison of effective temperature between this disk model and the one in W20. We consider the case of $M_\bullet=10^4~M_\odot$. The upper panel shows the temperature differences for different accretion rate for $a_\bullet=0.998$. The differences grow for lower accretion rate, because the radiative efficiency $\eta$ is bigger for low $\dot m$, where advection cooling is unimportant. The angular momentum removes by radiation is larger for a lower accretion rate disk. As a result, the lower accretion rate disk becomes dimmer. The lower panel shows the temperature differences with disk radius for different $a_\bullet$. Here we fix $\dot  m= 0.1$. The differences become bigger as $a_\bullet$ increases, again because $\eta$ becomes bigger as $a_\bullet$ increases. As a result, the radiation removes more angular momentum, making the disk dimmer. As we can see from both panels, for cases of high spin and low accretion, the effective temperature becomes dimmer by less than $10\%$.


\section{choice of outer and inner radius}
\label{rout_rin}
\begin{figure}[ht!]
\gridline{\fig{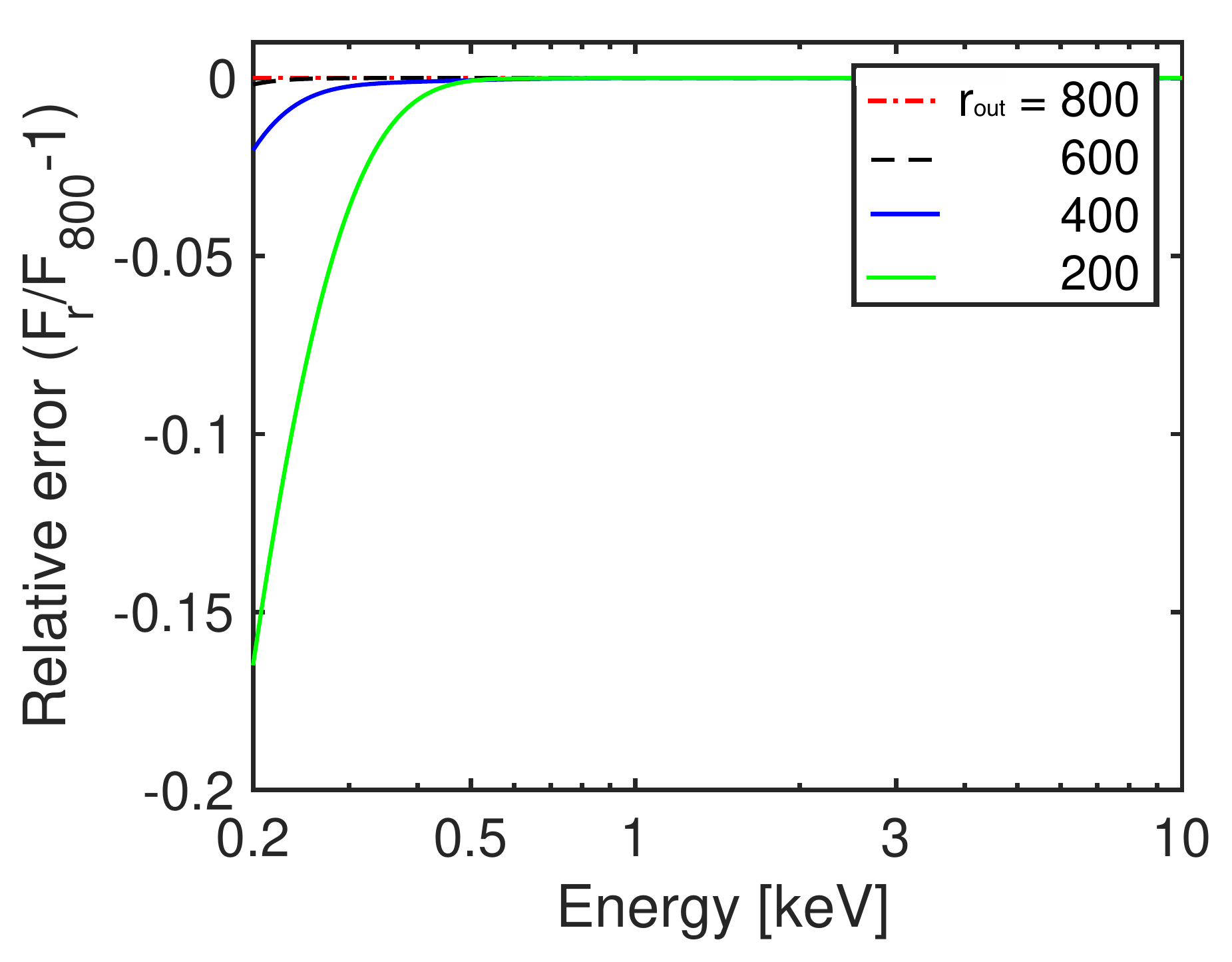}{0.45\textwidth}{}
          }
\caption{The effects on the spectrum of different choices of outer disk edge. The error is calculated as $E=\frac{F_{r}-F_{800}}{F_{800}}$ ($F_r$ denotes the flux of the disk with $r_{out}=r$) for a given frequency, and is $< 1.0\%$ when $r_{out}>600 ~\rm{R_g}$. For lower accretion rates, the error would be smaller due to a cooler disk.
}
\label{rout}
\end{figure}

\begin{figure}[ht!]
\plotone{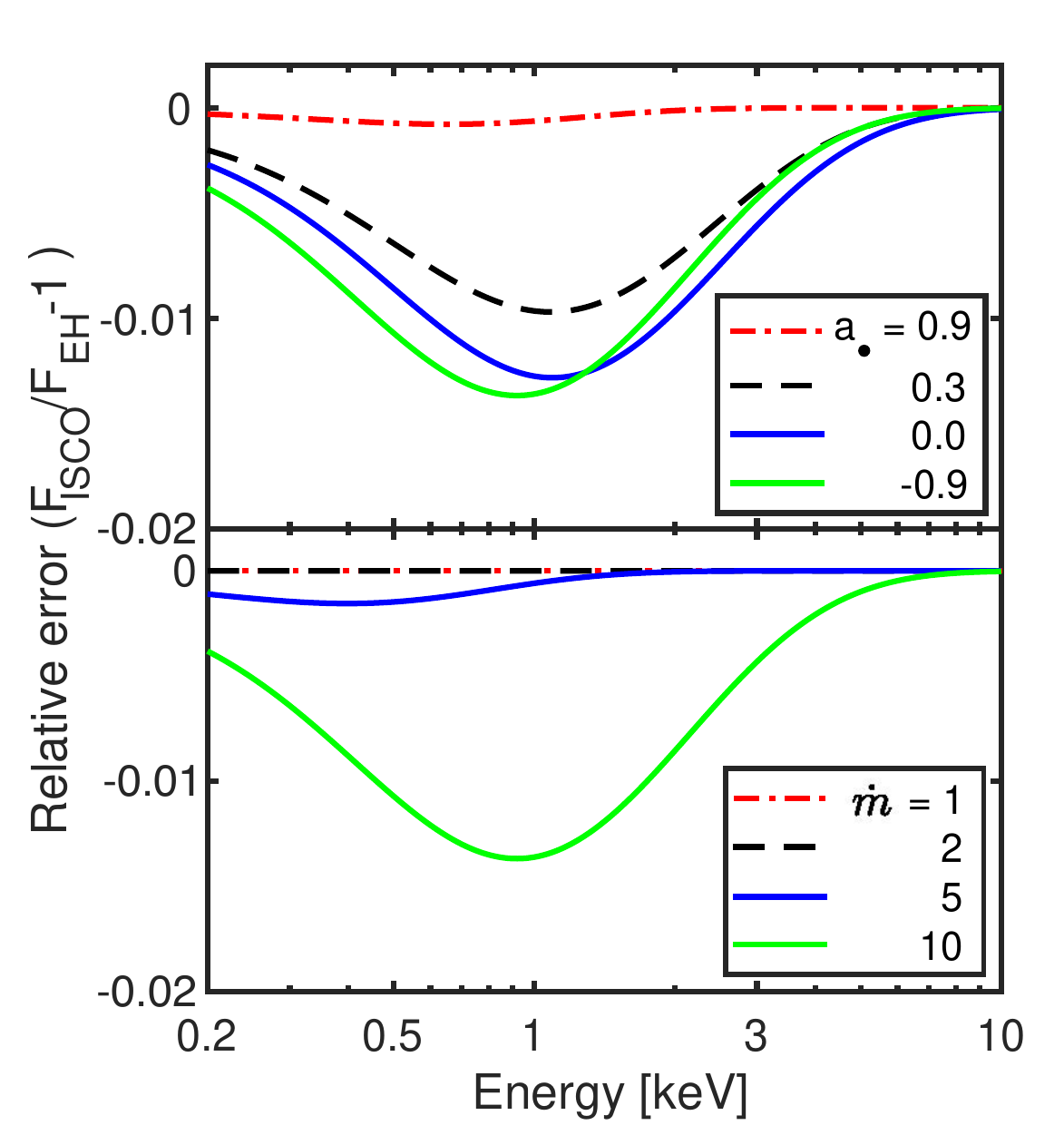}
\caption{The effects on the spectrum of different choices of inner disk edge for different $a_\bullet$ and $\dot m$ values. Here, two kinds of inner edge are considered: event horizon and ISCO. The error is calculated as $E=\frac{F_{EH}-F_{ISCO}}{F_{EH}}$ for a given frequency. 
We set $M_\bullet =10^4 M_\odot$ for both panels. For the upper panel, $\dot m = 10$ Edd, while for the lower panel $a_\bullet = -0.9$. This figure shows that the error is $< 2\%$ over a range of disk inner edges. 
}
\label{rin}
\end{figure}
In this section, we test the effects of different choices of disk outer and inner radius. Most of the X-ray photons are emitted in the inner disk (within  30~$R_g$, see W20). However, for an IMBH disk, the effective temperature at several hundred gravitational radii can be as high as several $\times 10^5 ~\rm{K}$. As a result, it could impact the low energy part of the 0.3--7 keV spectrum.   

Fig.~\ref{rout} shows the effects of different choices for the outer radius. We consider a disk with parameters $M_\bullet=10^4M_\odot$, $a_\bullet=0.998$, and $\dot m=10$ Edd. For this disk, two times the tidal radius is $2r_t=2000 ~\rm{R_g}$.
We fix the disk inclination at $45^\circ$, and calculate the spectrum for different outer disk radii. The relative error on the flux is $< 1 \%$ when $r_{out}>600 ~\rm{R_g}$. A lower $M_\bullet$, a lower $a_\bullet$ and a lower $\dot m$ produce a lower disk temperature in the outer disk region, moving the spectrum to a lower energy and making the relative error smaller.
Therefore, in the main paper we fix the outer radius at 600 $~\rm{R_g}$, even if the disk outer radius is bigger than 600$~\rm{R_g}$. The main effect is to accelerate the calculation, as the error introduced this way is negligible (Fig.~\ref{rout}).

Fig.~\ref{rin} shows the effect of the difference between using the ISCO or the event horizon as the inner disk radius for purposes of ray tracing. For all the disks considered here, we fix the inclination at $\theta=45^\circ$ and the BH mass at $M_\bullet=10^4M_\odot$. For high accretion rate disks, the slim disk assumptions would push the inner edge inside the ISCO. As a result, our choice to use the ISCO as the inner disk edge \citep{Sadowski09} could impact the spectrum and therefore the best-fit parameters. In Fig.~\ref{rin} we compare the results obtained if we set the boundary condition of the slim disk at close to the event horizon. In the main text, we cut off the disk at the ISCO when ray tracing, due to a singularity in the $f_c$ prescription. Instead, in this analysis when $f_c$ runs into this singularity, we reset it to 1. We study the slim disk model with different $a_\bullet$ and $\dot m$. For the upper panel, we fix the accretion rate at $\dot m=10$ Edd. The negative spin disk yields the biggest error, because the ISCO is relatively far away from the BH, and for a high accretion rate disk, the inner edge can be pushed to near the event horizon even for such a retrograde spinning disk. In the lower panel, we fix $a_\bullet=-0.9$. There is little difference between the spectra when choosing the ISCO or the event horizon as the inner edge for a low accretion rate disk. This is because for low accretion rate disks, the disk terminates at ISCO, and extending the disk inward would not affect the spectrum too much. For both panels, the error caused by choosing ISCO as inner radius is always $< 2\%$. For a lower accretion rate disk, the error is lower. As a result, we set the inner edge of the disk at ISCO in the main paper.

\end{appendix}

\end{document}